\documentclass[a4paper,12pt]{article}
\usepackage[utf8]{inputenc}
\usepackage[english]{babel}
\usepackage{graphicx}
\usepackage{mathptmx}
\usepackage[onehalfspacing]{setspace}
\usepackage[headheight=1in,margin=1in]{geometry}

\usepackage{amsmath, amssymb, amsthm}
\usepackage[authoryear,sort]{natbib}
\theoremstyle{definition}
\newtheorem{definition}{Definition}

\theoremstyle{plain}
\newtheorem{theorem}{Theorem}

\newtheorem{proposition}{Proposition}
\newtheorem{corollary}{Corollary}

\usepackage{titlesec}
\titlespacing{\paragraph}{0pt}{0pt}{1em}

\usepackage{subcaption}

\usepackage[bottom,hang,flushmargin]{footmisc}
\usepackage{xcolor}
\definecolor{sunset}{RGB}{238,93,108}
\definecolor{dusk}{RGB}{60,77,201}

\usepackage[
    hidelinks,
    unicode,
    bookmarksopen=true,
    bookmarksnumbered=true,
    linktoc=all
]{hyperref}

\graphicspath{{figures/}}

\usepackage{float}

\title{\vspace{-1in}\singlespacing Anytime-Valid Inference in Conjoint Experiments}

\author{Daniel Molitor\thanks{Department of Information Science, Cornell University, djm484@cornell.edu.} \and Jennah Gosciak\thanks{Department of Information Science, Cornell University, jrg377@cornell.edu.} \and Michael Lindon\thanks{Netflix, Los Gatos, CA, michael.s.lindon@gmail.com.}}
 
\date{}

\begin{document}

\maketitle

\begin{singlespacing}
\begin{abstract}
    Choosing an appropriate sample size is a fundamental challenge in experimental design. This challenge is particularly salient in conjoint experiments, where researchers simultaneously target many estimands whose inferential goals may be reached at different sample sizes. Conventional sample-size calculations address this challenge in an \emph{ex ante} manner. In large online studies, including many conjoint experiments, data often arrive continuously, allowing researchers to dynamically stop data collection. In such settings, the relevant practical question is not ``what sample size should I choose in advance?'' but rather, as data accumulate, ``have I learned enough yet?'' A naive response is to repeatedly inspect conventional confidence intervals and \(p\)-values, but doing so invalidates their inferential guarantees. We address this challenge by developing a general anytime-valid framework for conjoint experiments that allows continuous monitoring and data-dependent stopping. This framework equips many common conjoint estimands with asymptotic \(e\)-processes, confidence sequences, and anytime-valid $p$-values, and provides a general recipe for any estimand that can be expressed as a linear contrast of population OLS coefficients. Across common conjoint settings, we demonstrate that anytime-valid inference can yield substantial sample savings relative to fixed-sample designs. The probability of early stopping frequently exceeds 50\%, with mean sample savings often surpassing 25\%.
\end{abstract}
\end{singlespacing}

\noindent \textit{Keywords: conjoint experiments, anytime-valid inference, sequential decision-making, causal inference, power analysis}

\newpage

\section{Introduction}
\label{sec:introduction}

Choosing an appropriate sample size is a fundamental challenge in experimental
design. If a study collects too little data, substantively meaningful 
effects may remain too imprecisely estimated to support confident 
inference, and even statistically significant estimates may be exaggerated 
in magnitude or wrong in sign \citep{gelman_beyond_2014}. If a
study collects substantially more data than necessary, researchers may incur
unnecessary costs in resources, time, and effort. For this 
reason, sample-size planning is central to the design of credible and 
efficient empirical studies.

These concerns are especially relevant in conjoint experiments. Since the 
seminal work of \citet{hainmueller_causal_2014}, conjoint experiments 
have become a widely used tool for studying multidimensional preferences 
in political science and beyond.
A strength of conjoints is 
that they permit researchers to learn about many estimands simultaneously.
Yet this multi-dimensional structure can 
complicate design decisions. In many conjoints, the researcher is 
not interested in a single treatment effect, but in a collection of 
estimands, where estimates may differ substantially in both effect size and 
statistical precision. As a result, determining an appropriate sample 
size is often far from straightforward.

The standard method for determining an appropriate sample size is power analysis. Typically, a researcher specifies a minimum effect size of interest and 
then calculates the sample size required to detect that effect with a 
desired probability using a chosen test statistic. Recent work has 
made power analysis in conjoints more tractable for estimands such as average marginal component effects (AMCEs) and average component interaction effects (ACIEs) \citep{hainmueller_causal_2014, stefanelli_subjects_2020, schuessler_power_2020}. These 
works address an important but fundamentally \emph{ex ante} design problem: 
how to choose a fixed sample size before data collection begins.

In large online studies, however, data often arrive continuously and researchers may have substantial flexibility about whether to stop immediately or continue collecting additional data. 
In such settings, ex ante sample size calculations may be unnecessarily constraining and costly.
More specifically, conjoints frequently target many estimands simultaneously. Even when researchers have substantive expectations about the approximate size or direction of some effects, they may not know in advance which estimands will be estimated most precisely, which effects will be distinguishable from zero earliest, or whether the study's inferential goals will be met well before a pre-specified sample size is reached. In such settings, the relevant practical question is often not simply ``what fixed sample size should I choose in advance?'' but rather, as data accumulate, ``have I learned enough yet?''

A researcher who wishes to adapt data collection in response to the observed signal in the data may be tempted to repeatedly estimate treatment effects throughout the course of the experiment and decide to continue collecting data based on whether the results appear ``significant enough'' or ``precise enough,'' a practice often referred to as data peeking.
However, conventional confidence intervals and $p$-values do not retain their inferential guarantees under peeking, and repeated peeking can greatly inflate the probability of committing a Type I error \citep{simmons_false-positive_2011}. A natural goal, then, is to enable continuous monitoring in a statistically rigorous manner: to monitor accumulating evidence and adaptively make data-collection decisions without invalidating inferential guarantees.

To this end, we propose a new approach to conjoint analysis grounded in the anytime-valid inferential methods of \citet{lindon_clustered_2026}.
First, we equip a broad class of conjoint estimands with asymptotic \(e\)-processes, confidence sequences (CSs) and anytime-valid $p$-values. We then provide a general recipe that extends this approach to any estimand that can be expressed as a linear contrast of population OLS coefficients.
Second, we describe practical anytime-valid stopping rules based on statistical significance and estimation precision. Third, we demonstrate that this approach can yield substantial sample savings relative to fixed-sample designs, while avoiding the inferential invalidity induced by naive data peeking. Together, these contributions position anytime-valid inference as a practical and statistically principled tool for continuous monitoring in conjoint experiments.

Our contributions bridge two largely distinct bodies of literature. In the conjoint literature, prior work has focused on developing substantively interesting causal and descriptive estimands and estimating them with conventional regression methods \citep{hainmueller_causal_2014, leeper_measuring_2020, de_la_cuesta_improving_2022, egami_causal_2019}, as well as on developing power calculations for conjoint designs targeting those estimands \citep{schuessler_power_2020, stefanelli_subjects_2020}. In parallel, a growing methodological literature has developed anytime-valid inferential methods for conventional randomized experiments and related settings \citep{cook_semiparametric_2024, johari_always_2022, ham_design-based_2023, lindon_anytime-valid_2026, lindon_clustered_2026, molitor_anytime-valid_2025, liang_experimental_2024}. We merge these literatures by connecting anytime-valid inference to a broad class of common conjoint estimands, enabling valid continuous monitoring and data-driven stopping rules without sacrificing inferential guarantees. Our contributions also complement work on adaptivity in conjoint experiments. One strand of this literature studies adaptive treatment assignment to improve experimental efficiency \citep{offerwestort_adaptive_2021, gosciak_adaptive_2026}, while another adaptively personalizes conjoint tasks to improve preference elicitation \citep{yu_individually_2011, cunningham_adaptive_2010}. In contrast, we study adaptivity in data collection through continuous monitoring and stopping rules to improve sample efficiency and reduce operational costs.

The remainder of the paper proceeds as follows. Section~\ref{sec:savi} introduces the main tools of anytime-valid inference, including $e$-values, $e$-processes, confidence sequences, sequential $p$-values, and their asymptotic counterparts.
Section~\ref{sec:conjoint_review} reviews conjoint experiments, shows how common estimands map to linear contrasts of population OLS coefficients, and motivates the need for cluster-robust inference. Section~\ref{sec:av_inference} reviews conventional cluster-robust inference and anytime-valid cluster-robust inference.
Section~\ref{sec:av_estimation} formalizes a general recipe for anytime-valid inference on a broad class of common conjoint estimands.
Section~\ref{sec:results} uses simulations and a large asylum-seeker conjoint experiment to illustrate continuous monitoring, demonstrate the risks of naive peeking, and quantify the sample-efficiency gains of anytime-valid inference. Finally, Section~\ref{sec:discussion} summarizes the main results, discusses implications for conjoint design and analysis, and outlines directions for future research.

\section{Anytime-valid inference --- a primer}
\label{sec:savi}

Anytime-valid inference replaces the
fixed-sample guarantees of classical hypothesis testing and parameter estimation with guarantees that hold
simultaneously at every sample size, allowing for continuous monitoring
and data-dependent stopping without invalidating inference \citep{ramdas_game-theoretic_2023,
ramdas_hypothesis_2025}. Throughout this section, $t$
indexes the number of independent sampling units observed; in conjoint experiments, these units are respondents.

\subsection{Exact anytime-valid inference}
\label{sec:exact-av-inference}

\subsubsection{Hypothesis testing}
A central building block of anytime-valid inference is the \emph{$e$-value}, a measure of evidence against a null
hypothesis $H_0$, where $H_0$ is a set of distributions. An $e$-value is a nonnegative
random variable $E$ satisfying $\mathbb E_P\left[E\right]\leq1$ for every $P\in H_0$. 
Markov's
inequality yields an $\alpha$-level test: $\mathbb P_P(E\geq\alpha^{-1})\leq\alpha$ under every null
distribution, so $E\geq\alpha^{-1}$ provides sufficient evidence to reject \(H_0\) at level $\alpha$. The corresponding $p$-value is \(\min\left\{1, \frac{1}{E}\right\}\).

In many online experiments, data arrive sequentially. In these settings, we turn to \emph{\(e\)-processes}, the sequential counterpart of \(e\)-values.
An \emph{\(e\)-process} is a nonnegative sequence \((E_t)_{t\geq1}\) in which \(E_t\) depends only on the data observed through time \(t\), and where \(E_\tau\) is an \(e\)-value for every stopping time \(\tau\), where \(\tau\) can be chosen in a data-driven manner.
By Ville's inequality, under \(H_0\), the probability that the \(e\)-process ever reaches the rejection threshold \(\alpha^{-1}\) is  at most \(\alpha\) \citep{ramdas_hypothesis_2025}:
\begin{equation}
\mathbb P_P\left\{E_t\geq\alpha^{-1}\ \text{for some }t\geq1\right\}\leq\alpha
\qquad\text{for every }P\in H_0.
\label{eq:ville}
\end{equation}
This result carries a powerful implication: a researcher may reject \(H_0\) the first time \(E_t\geq\alpha^{-1}\) while controlling Type I error at level \(\alpha\), regardless of how often the researcher examines the results or what data-dependent stopping rule they use.

\subsubsection{Parameter estimation} 
Beyond hypothesis testing, \(e\)-processes also 
yield familiar inferential objects: \emph{confidence sequences} and anytime-valid \emph{p-values}.

A sequence of sets $(C_t)_{t\geq1}$ is a $1-\alpha$ \emph{confidence sequence} (CS) for a parameter $\theta$ if
\begin{equation}
\mathbb P\{\theta\in C_t\ \text{for all }t\geq1\}\geq1-\alpha.
\label{eq:cs-guarantee}
\end{equation}
Unlike a conventional confidence interval (CI), whose coverage guarantee applies at a single prespecified sample size, a CS provides simultaneous coverage across all sample sizes. Researchers can therefore monitor the CS as data accrue and stop data collection based on the observed results without invalidating its coverage guarantee \citep{howard_time-uniform_2021}.

A CS can be constructed by inverting a family of anytime-valid hypothesis tests, one for each possible value of the parameter of interest. 
Suppose that, for every candidate value \(\theta_0\), \(E_t(\theta_0)\) is an \(e\)-process for testing \(H_0:\theta=\theta_0\). 
At each time \(t\), the CS retains every candidate value that has not been rejected:
\[
C_t
=
\left\{
\theta_0:E_t(\theta_0)<\alpha^{-1}
\right\}.
\]
By \eqref{eq:ville}, the probability that the true value of \(\theta\) is excluded at any point is at most \(\alpha\), giving the simultaneous coverage guarantee of \eqref{eq:cs-guarantee}.

The same \(e\)-process yields an \emph{anytime-valid \(p\)-value} by setting \(p_t=\min\{1,E_t^{-1}\}\) at each time \(t\). An anytime-valid \(p\)-value is a sequence \((p_t)_{t\geq1}\) satisfying, under \(H_0\),
\[
\mathbb P\left(p_t\leq\alpha\ \text{for some }t\geq1\right)\leq\alpha
\]
for every \(\alpha\in(0,1)\). It can therefore be interpreted like a conventional \(p\)-value while remaining valid under continuous monitoring and data-dependent stopping.

\subsection{Asymptotic anytime-valid inference}

Exact \(e\)-processes are easiest to construct when \(H_0\) imposes additional structure, such as a parametric likelihood or known bounds on the relevant random variable. For example, under a simple null hypothesis, the likelihood ratio comparing the null against a fixed alternative can be updated as data accumulate. The resulting sequence is an \(e\)-process, so \eqref{eq:ville} applies directly.

However, many estimands of interest are defined without any such structure. A population least-squares regression coefficient, for example, describes the best linear prediction of an outcome from a set of regressors without requiring a complete model of the outcome distribution.
In these settings, researchers can use asymptotic \(e\)-processes.
Asymptotic \(e\)-processes play the same role as the central limit theorem in conventional inference: they replace exact finite-sample guarantees with approximations that become increasingly accurate as data accumulate.

Formally, let \(\mathcal P\) denote the class of possible data-generating distributions and \(H_0\subset\mathcal P\). A nonnegative process \((E_t)_{t\geq1}\) is an \emph{asymptotic \(e\)-process} for \(H_0\) if there exists an \(e\)-process \((E_t^S)_{t\geq0}\) for \(H_0\), possibly defined on an enriched probability space, such that
\begin{equation}
\frac{\log E_t}{\log E_t^S}\to1
\qquad\text{almost surely under every \(P\in\mathcal P\)}
\label{eq:asymptotic_eprocess}
\end{equation}
\citep[Definition~C.2]{waudby-smith_time-uniform_2024}.\footnote{The enriched space and comparison \(e\)-process may depend on \(P\).} Intuitively, the asymptotic \(e\)-process \(E_t\) becomes equivalent on the log scale to an exact \(e\)-process \(E_t^S\) as the sample grows. From this asymptotic \(e\)-process, we can construct an asymptotic CS,
\[
C_t
=
\left\{
\theta_0:E_t(\theta_0)<\alpha^{-1}
\right\},
\]
and an asymptotic anytime-valid \(p\)-value,
\(p_t=\min\{1,E_t^{-1}\},\)
in the same manner as in Section~\ref{sec:exact-av-inference}.

\subsection{Anytime-valid stopping rules}

These inferential tools allow researchers to continuously monitor their estimates and stop data collection whenever they choose. However, researchers may prefer to use a data-driven stopping rule to determine when they have collected enough data to meet their inferential goals. We highlight three practical options below.

A \emph{significance-based stopping rule} stops data collection when the anytime-valid \(p\)-values for one or more focal estimates fall below a chosen significance level \(\alpha\) \citep{johari_always_2022}. This rule is most natural when the researcher expects that the focal effects are non-zero but is uncertain about their magnitude. Significance-based stopping allows the study to end once the data have provided sufficient evidence that the focal effects are distinguishable from \(0\).

A \emph{precision-based stopping rule} stops once the CSs for the focal estimates reach a maximum width, regardless of whether they exclude \(0\) \citep{frey_fixed-width_2010}. This rule is particularly useful when the research goal is not merely to establish that an effect is non-zero, but to estimate it with sufficient precision. Precision-based stopping also provides a principled way to ensure that null or near-null effects are estimated precisely.

A \emph{hybrid stopping rule} combines the previous stopping rules. For example, a researcher may choose to stop once the focal estimates are both distinguishable from \(0\) and estimated with sufficient precision. This avoids stopping when an effect first becomes marginally statistically significant but remains imprecisely estimated.

\section{Conjoint estimands and cluster-robust inference}
\label{sec:conjoint_review}

Conjoint experiments are randomized experiments designed to study multidimensional human preferences. In a typical forced-choice conjoint, each respondent completes one or more choice tasks. In each task, the respondent is asked to choose between two or more profiles, such as political candidates or job applicants, where each profile is described by a vector of attributes. Attributes are generally categorical variables, and their possible values are called levels.
For example, a political candidate profile might include attributes for political party, gender, education, and geographic region, each taking one of several possible levels.
The researcher generates each profile by jointly randomizing its attributes according to a prespecified distribution over the attribute space. This randomization enables the identification and estimation of a variety of causal and descriptive estimands characterizing respondents' preferences.

We focus on forced-choice designs for concreteness, but the following notation and discussion also extends to ratings, rankings, and other outcome structures. Let \(i=1,\ldots,G\) index respondents, \(k\) index tasks, \(m\) index profiles within a task, and \(j=1,\ldots,J\) index attributes. Let \(\mathcal{L}_j\) denote the set of possible levels of attribute \(j\), and let \(\mathcal{A}=\prod_{j=1}^{J}\mathcal{L}_j\) denote the profile space consisting of all possible attribute vectors. Let \(A_{ikm}=(A_{ikm,1},\ldots,A_{ikm,J})\in\mathcal{A}\) denote the realized attribute vector for profile \(m\) in task \(k\) for respondent \(i\), and let \(Y_{ikm}\) denote the observed profile-level outcome.
\[
    Y_{ikm} = 
    \begin{cases}
    1 & \text{if respondent chooses profile } m \\
    0   & \text{otherwise}.\\
\end{cases}
\]

\subsection{The average marginal component effect}
\label{sec:amce}

One of the most commonly targeted causal conjoint estimands is the average marginal component effect (AMCE), introduced by \citet{hainmueller_causal_2014}. Let \(Y_i(a)\) denote respondent \(i\)'s potential outcome for a profile with attribute vector \(a\). Let \(\bar Y(a)=E[Y_i(a)]\) denote the corresponding population mean potential outcome, where the expectation averages over respondents and the randomized attributes of the competing profiles in the task.

Fix an attribute \(j\), a target level \(\ell\in\mathcal{L}_j\), and a reference level \(\ell_{0}\in\mathcal{L}_j\). For any attribute vector \(a=(a_1,\ldots,a_J)\in\mathcal{A}\), let \(a_{-j}=(a_1,\ldots,a_{j-1},a_{j+1},\ldots,a_J)\) denote the vector containing all attributes other than \(j\). Accordingly, let \(A_{-j}\) denote the random vector of all non-focal attributes and let \(\pi_{-j}\) denote its joint distribution. We use \(\bar Y(\ell,a_{-j})\) to denote the population mean potential outcome when attribute \(j\) is set to \(\ell\) and the remaining attributes are set to \(a_{-j}\).

\begin{definition}[Average marginal component effect]
\label{def:amce}
The AMCE of level \(\ell\) relative to \(\ell_{0}\) is
\[
\tau_{j_\ell}
=
E_{A_{-j}\sim\pi_{-j}}
\left[
\bar Y(\ell,A_{-j})
-
\bar Y(\ell_{0},A_{-j})
\right].
\]
\end{definition}

\noindent The AMCE is the average change in the probability that a profile is chosen when attribute \(j\) is changed from \(\ell_{0}\) to \(\ell\), holding all other attributes fixed and then averaging over their joint distribution \(\pi_{-j}\).\footnote{Adopting the convention \(\tau_{j_{\ell_{0}}}=0\), the comparison between any two levels \(\ell,\ell'\in\mathcal{L}_j\) is the linear contrast \(\tau_{j_\ell}-\tau_{j_{\ell'}}=E_{A_{-j}\sim\pi_{-j}}[\bar Y(\ell,A_{-j})-\bar Y(\ell',A_{-j})]\). A comparison between two non-reference levels requires no change of baseline, and contrasts of this form fall within the scope of the inferential framework developed in Section~\ref{sec:av_inference}.}

When the standard assumptions of consistency, no carryover effects across tasks, and no profile-order effects hold, observed outcomes can be linked to the relevant potential outcomes \citep{hainmueller_causal_2014}. Under complete independent randomization (\citealp{hainmueller_causal_2014}, Proposition~3),
\[
E\left[Y_{ikm}\mid A_{ikm,j}=\ell\right]
=
E_{A_{-j}\sim\pi_{-j}}
\left[
\bar Y(\ell,A_{-j})
\right],
\]
and consequently,
\[
\tau_{j_\ell}
=
E\left[Y_{ikm}\mid A_{ikm,j}=\ell\right]
-
E\left[Y_{ikm}\mid A_{ikm,j}=\ell_{0}\right].
\]
Thus, the AMCE is identified by the difference in observed mean outcomes between profiles assigned the target level and profiles assigned the reference level.

\subsection{OLS estimation and cluster-robust inference}
\label{sec:population_projection}

The identification result in Section~\ref{sec:amce} motivates the standard OLS estimator. We first define the population quantity that OLS estimates, which provides the link between conjoint estimands and the general anytime-valid inferential framework in Section~\ref{sec:av_inference}.

\begin{definition}[Population least-squares projection]
\label{def:projection}
Let \((Y,X)\sim P\) denote a generic sampling unit, where \(Y\) is an outcome vector and \(X\) is a \(p\)-column design matrix with the same number of rows. The population least-squares projection of \(Y\) onto \(X\) is \(X\beta^\star\), where
\begin{equation}
\beta^\star
=
\arg\min_{\beta\in\mathbb{R}^p}
\mathbb E_P\!\left[(Y-X\beta)'(Y-X\beta)\right]
=
\arg\min_{\beta\in\mathbb{R}^p}
\mathbb E_P\!\left[\|Y-X\beta\|_2^2\right].
\label{eq:population_projection}
\end{equation}
\end{definition}

If the required second moments exist and \(\mathbb E_P\left[X'X\right]\) is positive definite, the coefficient vector \(\beta^\star\) is unique. The projection \(X\beta^\star\) provides the best linear prediction of the outcome from the included regressors under squared loss, without requiring the conditional mean \(\mathbb E_P\left[Y\mid X\right]\) to be linear. Writing the outcome as \(Y=X\beta^\star+\varepsilon^\star\), where \(\varepsilon^\star=Y-X\beta^\star\), gives the corresponding population least-squares regression.

Let \(Y_i\) collect the profile-level outcomes contributed by respondent \(i\), and let \(X_i\) contain the corresponding regressors. For every non-reference level \(\ell\in\mathcal{L}_j\setminus\{\ell_{0}\}\) of attribute \(j\), define \(D_{ikm,j_\ell}=\mathbf{1}\{A_{ikm,j}=\ell\}\). Using an intercept and the indicators for all non-reference attribute levels, the population least-squares regression can be written as
\[
Y_{ikm}
=
\alpha^\star
+
\sum_{j=1}^{J}
\sum_{\ell\in\mathcal{L}_j\setminus\{\ell_{0}\}}
\beta_{j_\ell}^\star D_{ikm,j_\ell}
+
\varepsilon_{ikm}^\star.
\]
Here \(\beta^\star=(\alpha^\star,(\beta_{j_\ell}^\star)_{j,\ell})\) is the population coefficient vector from Definition~\ref{def:projection}, with \(X_i\) formed by stacking the rows \((1,(D_{ikm,j_\ell})_{j,\ell})\). 
Because the number of profile evaluations completed by each respondent is independent of the profile assignments and potential outcomes, 
the respondent-level and profile-level least-squares objectives differ only by a positive constant and therefore have the same minimizer.

\begin{proposition}[AMCEs as population least-squares coefficients; \citealp{hainmueller_causal_2014}, Proposition~3]
\label{prop:amce_projection}
Suppose the identification and randomization conditions described in Section~\ref{sec:amce} hold. Then, for every attribute \(j\) and non-reference level \(\ell\in\mathcal{L}_j\setminus\{\ell_{0}\}\),
\(
\beta_{j_\ell}^\star
=
\tau_{j_\ell}.
\)
\end{proposition}

\noindent Appendix~\ref{appendix:amce_projection_proof} provides a proof.

Proposition~\ref{prop:amce_projection} establishes that the AMCE \(\tau_{j_\ell}\) can be represented by the population least-squares regression coefficient \(\beta_{j_\ell}^\star\). The corresponding sample OLS coefficient \(\widehat{\beta}_{j_\ell}\) is therefore the usual estimator of the AMCE \citep{hainmueller_causal_2014}. This representation does not require additive or linear preferences: independent randomization of the attributes ensures that the population regression coefficient equals the AMCE even when potential outcomes depend on nonlinear relationships or interactions among attributes.

Although the profile attributes are randomized, the profile-level observations are not generally independent. Each respondent typically completes multiple tasks and evaluates multiple profiles within each task, inducing dependence among the outcomes contributed by that respondent. As a result, inference that treats every profile-level observation as independent will generally understate uncertainty. Conjoint analysis addresses this dependence using standard errors clustered at the respondent level, allowing outcomes from the same respondent to be arbitrarily dependent while treating respondents as independent sampling units \citep{hainmueller_causal_2014}.

\subsection{A general recipe for conjoint inference}
\label{sec:conjoint_estimand_recipe}

The AMCE illustrates a general strategy for estimating common conjoint estimands. First, the researcher defines an estimand of interest. Second, the researcher establishes that the estimand is nonparametrically identified under the randomized conjoint design and relevant identification assumptions. Finally, the researcher represents the estimand as a coefficient or linear contrast from an appropriately specified population least-squares regression and performs inference using the corresponding sample regression estimator with respondent-clustered standard errors.

Many common conjoint estimands follow this recipe. AMCEs, conditional AMCEs, ACIEs, population AMCEs (pAMCEs), average marginal interaction effects (AMIEs), and marginal means are nonparametrically identified under standard assumptions such as independently randomized profile attributes and consistency \citep{hainmueller_causal_2014, egami_causal_2019, de_la_cuesta_improving_2022, leeper_measuring_2020}. Moreover, each can be represented as a coefficient or linear contrast from an appropriately specified population least-squares regression. Different estimands may require different regression specifications, such as interactions between attributes or between attributes and respondent characteristics.

Consequently, anytime-valid inference for a broad class of conjoint estimands can be reduced to anytime-valid inference for linear contrasts of population least-squares regression coefficients. The central challenge is that each independent respondent may contribute multiple dependent outcomes, requiring inference to account for clustering. Section~\ref{sec:av_inference} presents the cluster-robust anytime-valid inference methods needed to address this problem.

\section{Cluster-robust anytime-valid inference for linear models}
\label{sec:av_inference}

This section introduces anytime-valid inference for linear contrasts of population least-squares regression coefficients when each independent sampling unit contributes multiple dependent outcomes. We first review conventional cluster-robust inference and then state the corresponding anytime-valid result.
Although our application centers on conjoint experiments, the methodology applies more generally to any setting that requires cluster-robust least-squares regression.

\subsection{Conventional cluster-robust inference}
\label{sec:fixed_g_reference}

Let \(g=1,\ldots,G\) index clusters, and let \(n_g\) denote the number of observations in cluster \(g\). Cluster \(g\) has outcome vector \(Y_g\in\mathbb{R}^{n_g}\) and design matrix \(X_g\in\mathbb{R}^{n_g\times p}\). Outcomes may be arbitrarily dependent within each cluster, but clusters are independent. We retain the population least-squares coefficient vector \(\beta^\star\) defined in Equation~\eqref{eq:population_projection}. Based on the first \(G\) clusters, its sample OLS estimator is
\begin{equation}
\widehat\beta_G
=
A_G^{-1}\sum_{g=1}^G X_g'Y_g,
\qquad
\text{where \quad} A_G=\sum_{g=1}^G X_g'X_g.
\label{eq:cluster_ols_estimator}
\end{equation}
Let \(\widehat u_g=Y_g-X_g\widehat\beta_G\) denote the residual vector for cluster \(g\), and define its estimated score as \(\widehat s_g=X_g'\widehat u_g\). The cluster-robust covariance estimator is
\begin{equation}
\widehat V_G
=
\frac{G}{G-1}\,A_G^{-1}\widehat M_GA_G^{-1},
\qquad
\widehat M_G
=
\sum_{g=1}^G\widehat s_g\widehat s_g',
\label{eq:cluster_covariance_estimator}
\end{equation}
where \(G/(G-1)\) is the standard finite-cluster correction (CR1).
The estimated score \(\widehat s_g\) is the sample counterpart of the population cluster score
\begin{equation}
\psi_g
=
X_g'\left(Y_g-X_g\beta^\star\right).
\label{eq:population_score}
\end{equation}
Because they do not depend on the sample estimator \(\widehat\beta_G\), the population scores \(\psi_g\) are independent across clusters and form the basis of the anytime-valid result introduced in Section~\ref{sec:cluster_av_linear_models}.

Consider a \(q\)-dimensional linear contrast \(\theta=R\beta^\star\), where \(R\in\mathbb{R}^{q\times p}\) has full row rank. Conventional inference for \(H_0:R\beta^\star=\theta_0\) uses the cluster-robust Wald statistic
\begin{equation}
Q_G(\theta_0)
=
(R\widehat\beta_G-\theta_0)'
(R\widehat V_GR')^{-1}
(R\widehat\beta_G-\theta_0).
\label{eq:cluster_wald_stat}
\end{equation}
Let \(c_\alpha\) denote the chosen \(1-\alpha\) critical value. The conventional test rejects \(H_0\) when \(Q_G(\theta_0)>c_\alpha\), and inverting this test gives the confidence region
\begin{equation}
C_{G,\mathrm{fixed}}^\alpha
=
\left\{
\theta\in\mathbb{R}^q:
Q_G(\theta)\leq c_\alpha
\right\}.
\label{eq:fixed_g_confidence_region}
\end{equation}
In the scalar case \(q=1\), the standard choice \(c_\alpha=t_{G-1,1-\alpha/2}^2\) gives the familiar cluster-robust confidence interval
\begin{equation}
R\widehat\beta_G
\pm
t_{G-1,1-\alpha/2}
\sqrt{R\widehat V_GR'},
\label{eq:conventional-clust-robust-ci}
\end{equation}
\noindent where \(t_{G-1,1-\alpha/2}\) is the \(1-\alpha/2\) quantile of the \(t\)-distribution with \(G-1\) degrees of freedom.
Reference distributions for \(c_\alpha\), finite-cluster corrections, and their limitations are reviewed in Appendix~\ref{appendix:conventional_cluster_inference}.

The inferential guarantees of this conventional procedure apply at a fixed sample size \(G\) chosen independently of the observed results. It does not permit continuous monitoring and data-dependent stopping. Under continuous monitoring, the probability of ever rejecting a true null can substantially exceed \(\alpha\).

\subsection{Anytime-valid cluster-robust inference}
\label{sec:cluster_av_linear_models}

We now transform the Wald statistic in Equation~\eqref{eq:cluster_wald_stat} into an asymptotic anytime-valid inferential object using Theorem~2 of \citet{lindon_clustered_2026}, which applies a multivariate strong Gaussian approximation to the sequence of independent population cluster scores \(\psi_g\) defined in Equation~\eqref{eq:population_score}.

The result requires i.i.d.\ clusters, a stable and nondegenerate design matrix and contrast covariance, consistent estimation of the cluster-score covariance, and a multivariate strong Gaussian approximation to the cumulative cluster scores \(\sum_{g=1}^G\psi_g\).
These conditions are stated formally in \citet[Assumption~1]{lindon_clustered_2026}. Theorem~\ref{thm:conjoint_estimands} provides sufficient conditions for conjoint experiments.

\begin{theorem}[Cluster-robust asymptotic \(e\)-process; {\citealp[Theorem~2]{lindon_clustered_2026}}]
\label{thm:cluster_robust_e_process}
Suppose the cluster-score regularity conditions of \citet[Assumption~1]{lindon_clustered_2026} hold.\footnote{\(X_g\) and \(\beta^\star\) correspond to \(W_g\) and \(\gamma^\star\) in \citet{lindon_clustered_2026}, Theorem~2.} For a fixed tuning parameter \(\lambda>0\), let \(\nu_G=G-1\), where \(q\) is the number of rows of the contrast matrix \(R\in\mathbb{R}^{q\times p}\). The process
\begin{equation}
E_G(\theta_0)
=
\left(\frac{\lambda}{\lambda+G}\right)^{q/2}
\left(
\frac{
1+\frac{\lambda}{\lambda+G}\frac{Q_G(\theta_0)}{\nu_G}
}{
1+\frac{Q_G(\theta_0)}{\nu_G}
}
\right)^{-(\nu_G+1)/2},
\qquad G>q,
\label{eq:cluster_t_e_process}
\end{equation}
is an asymptotic \(e\)-process for \(H_0:R\beta^\star=\theta_0\). Set \(E_G(\theta_0)=1\) whenever \(G\leq q\), the OLS estimator is undefined, or \(R\widehat V_GR'\) is singular.
\end{theorem}

\noindent The proof is provided in \citet[Appendix~C]{lindon_clustered_2026}.

This construction uses the same OLS estimator, covariance estimator, and Wald statistic as conventional cluster-robust inference. It transforms \(Q_G(\theta_0)\) using \(\nu_G=G-1\) whether testing a single contrast or several contrasts jointly. In the scalar case, \(Q_G(\theta_0)\) is the squared cluster-robust \(t\)-statistic. Computing \(E_G(\theta_0)\) therefore requires only a simple, closed-form transformation of the conventional test statistic.

The process \(E_G(\theta_0)\) measures the evidence against \(H_0:R\beta^\star=\theta_0\), with larger values indicating stronger evidence. Researchers may monitor this evidence as clusters accrue and reject \(H_0\) when \(E_G(\theta_0) \geq \alpha^{-1}\).

\begin{corollary}[Asymptotic anytime-valid \(p\)-values; {\citealp[Corollary~4]{lindon_clustered_2026}}]
\label{cor:av_pvalue}
Under the conditions of Theorem~\ref{thm:cluster_robust_e_process},
\begin{equation}
p_G(\theta_0)
=
\min\left\{1,\frac{1}{E_G(\theta_0)}\right\},
\qquad G>q,
\label{eq:av_pval}
\end{equation}
is an asymptotic anytime-valid \(p\)-value for \(H_0:R\beta^\star=\theta_0\).\footnote{The untruncated reciprocal \(1/E_G(\theta_0)\) inherits the logarithmic equivalence in Definition~\ref{eq:asymptotic_eprocess}; truncation at 1 does not affect rejection decisions for \(\alpha\in(0,1)\).}
\end{corollary}

\noindent Inverting the same asymptotic \(e\)-process over possible values of \(\theta\) produces an asymptotic \(1-\alpha\) CS.

\begin{corollary}[Asymptotic cluster-robust confidence sequence; {\citealp[Corollary~5]{lindon_clustered_2026}}]
\label{cor:cluster_cs}
Under the conditions of Theorem~\ref{thm:cluster_robust_e_process},
\begin{equation}
C_G^\alpha
=
\left\{
\theta\in\mathbb{R}^q:
E_G(\theta)<\alpha^{-1}
\right\},
\qquad G>q,
\label{eq:cluster_cs_region}
\end{equation}
is an asymptotic \(1-\alpha\) confidence sequence for \(R\beta^\star\). Define
\[
u_G(\lambda,q,\alpha)
=
\left(
\frac{\alpha^{2/q}\lambda}{\lambda+G}
\right)^{q/(\nu_G+1)}
\]
and
\[
\kappa_G(\lambda,q,\alpha)
=
\frac{\nu_G}{q}
\frac{1-u_G(\lambda,q,\alpha)}
{\max\left\{0,\;
u_G(\lambda,q,\alpha)-\frac{\lambda}{\lambda+G}
\right\}}.
\]
Then \(C_G^\alpha\) is the open Wald ellipsoid
\[
C_G^\alpha
=
\left\{
\theta\in\mathbb{R}^q:
\frac{Q_G(\theta)}{q}
<
\kappa_G(\lambda,q,\alpha)
\right\}.\footnote{Under the singular-sample convention of \citet[Theorem~2]{lindon_clustered_2026}, or when the denominator defining \(\kappa_G(\lambda,q,\alpha)\) vanishes, \(C_G^\alpha=\mathbb R^q\). In the latter case, \(\kappa_G(\lambda,q,\alpha)=+\infty\).}
\]
For a scalar contrast \((q=1)\) this gives the open interval
\begin{equation}
C_G^\alpha
=
R\widehat\beta_G
\ \pm\
\sqrt{R\widehat V_GR'}\,
\sqrt{\kappa_G(\lambda,1,\alpha)}.
\label{eq:cluster_scalar_cs}
\end{equation}
\end{corollary}

\noindent Equation~\ref{eq:cluster_scalar_cs} has the same form as a conventional cluster-robust confidence interval \eqref{eq:conventional-clust-robust-ci}, but replaces the fixed-sample critical value with the sample-size-dependent factor \(\sqrt{\kappa_G(\lambda,1,\alpha)}\).

The tuning parameter \(\lambda\) determines what sample size \(G\) the CS and anytime-valid \(p\)-value are optimized for. Smaller values target smaller sample sizes, yielding more precise CSs and \(p\)-values earlier in data collection, while larger values target larger sample sizes.
If \(\lambda\) targets a planned sample size \(G^\star\), but the study reaches its inferential goal either before or after \(G^\star\), the resulting CS and \(p\)-value will be somewhat less precise than those optimized for the realized sample size.
In this sense, \(\lambda\) is analogous to the planned sample size in a conventional fixed-sample design.
A natural default, which we use in Section~\ref{sec:results}, is to select the value of \(\lambda\) that minimizes the CS radius at \(G=G^\star\). This makes the CS and \(p\)-value as precise as possible at the sample size at which a fixed-sample study would have ended. It is important to note that \(\lambda\) must be selected before continuous monitoring begins. Choosing it in response to the observed data forfeits the CS and anytime-valid \(p\)-value's inferential guarantees.

Anytime-valid inference requires minimal change to standard least-squares regression workflows. Researchers fit the same OLS regression and compute the same cluster-robust covariance estimator, but use the resulting estimate and standard error to construct a cluster-robust CS and anytime-valid \(p\)-value rather than a conventional cluster-robust CI and \(p\)-value. Next, we use this result to obtain anytime-valid inference for a broad class of common conjoint estimands.

\section{Anytime-valid estimation of conjoint estimands}
\label{sec:av_estimation}

\subsection{A general recipe for anytime-valid conjoint inference}
\label{sec:av_estimation_recipe}

As outlined in Section~\ref{sec:conjoint_estimand_recipe}, a common strategy for estimating target conjoint estimands proceeds in three steps. First, the researcher defines an estimand of interest, such as an AMCE. Next, the researcher shows that the estimand is nonparametrically identified under the conjoint design and relevant identification assumptions. Finally, the researcher shows that the estimand can be represented as a coefficient or linear contrast from an appropriately specified population least-squares regression.

This observation yields a general recipe for anytime-valid inference in conjoint experiments. For the AMCE, Proposition~\ref{prop:amce_projection} establishes that \(\tau_{j_\ell}\) equals the corresponding population regression coefficient \(\beta_{j_\ell}^\star\). Researchers can therefore fit the OLS regression described in Section~\ref{sec:population_projection}, compute the cluster-robust covariance estimator, and use the resulting coefficient estimate and standard error to construct the \(e\)-process, anytime-valid \(p\)-value, and CS defined in Section~\ref{sec:cluster_av_linear_models}. This same process applies to \textit{any} nonparametrically identified conjoint estimand that can be represented as a coefficient or linear contrast from a population least-squares regression. The following theorem formalizes this argument for conjoint estimands with such a representation.

\begin{theorem}[Asymptotic anytime-valid inference for conjoint estimands]
\label{thm:conjoint_estimands}
Suppose respondent clusters \((Y_i,X_i)\) are i.i.d., with a uniformly bounded number of profile evaluations per respondent and uniformly bounded regressors and outcomes. Additionally, suppose the target conjoint estimand is nonparametrically identified and can be represented as \(\theta=R\beta^\star\), where \(R\in\mathbb R^{q\times p}\) is fixed and has full row rank. Define
\(
\Sigma_R=RH^{-1}\Omega H^{-1}R'
\),
where
\(
H=\mathbb E(X_i'X_i)
\),
\(
\psi_i=X_i'(Y_i-X_i\beta^\star)
\),
and
\(
\Omega=\mathbb E(\psi_i\psi_i')
\).
Assume \(H\succ0\) and \(\Sigma_R\succ0\), where \(\succ0\) denotes positive definiteness. Then, applying Theorem~\ref{thm:cluster_robust_e_process} and Corollaries~\ref{cor:av_pvalue}--\ref{cor:cluster_cs} to the conventional OLS estimator \(\widehat\beta_G\) \eqref{eq:cluster_ols_estimator} and cluster-robust covariance estimator \(\widehat V_G\) \eqref{eq:cluster_covariance_estimator} yields an asymptotic \(e\)-process, anytime-valid \(p\)-value, and confidence sequence for \(\theta\).
\end{theorem}

\noindent Appendix~\ref{appendix:conjoint_theorem} provides a proof.

In standard conjoint designs, bounded cluster sizes, dummy-coded regressors, and binary (or bounded) outcomes satisfy the boundedness requirements of Theorem~\ref{thm:conjoint_estimands}. Thus, under mild assumptions, researchers can replace conventional fixed-sample inference with its asymptotic anytime-valid analogue for many common conjoint estimands.

\subsection{Multiple hypothesis testing}
\label{sec:multiple_hypothesis_testing}

A distinct concern in this setting is multiple hypothesis testing. Conjoint experiments are inherently high-dimensional, and researchers typically estimate many effects simultaneously. As \citet{liu_multiple_2023} emphasize, uncorrected inference can increase the risk of false discoveries and misleading coverage statements. Under continuous monitoring, these concerns can become even more acute. The ability to repeatedly inspect evolving estimates creates opportunities for researchers to focus on spurious patterns and, intentionally or not, allow those patterns to shape stopping decisions.

To address these concerns, practitioners can pair anytime-valid CSs and \(p\)-values with appropriate multiple-testing adjustments. For exact anytime-valid \(p\)-values, Bonferroni adjustment controls familywise error by a union bound \citep{dunn_multiple_1961}. Other familiar corrections require their usual validity and dependence conditions \citep{benjamini_controlling_1995, benjamini_false_2005}. In addition, specialized procedures have been developed for multiple testing and interval estimation using exact \(e\)-values. The \(e\)-BH procedure controls the false discovery rate under arbitrary dependence among \(e\)-values \citep{wang_false_2022}, a guarantee that the conventional Benjamini--Hochberg procedure does not provide without additional assumptions on the dependence among \(p\)-values. For interval estimation, the \(e\)-BY procedure controls the false coverage rate for selected intervals constructed by inverting \(e\)-values \citep{xu_post-selection_2024}.

\section{Empirical results}
\label{sec:results}

In this section, we provide four simulation studies and one real-data case study of our proposed framework. Throughout, we focus on AMCEs because they are one of the most common conjoint estimands; however, all results apply directly to the broader class of estimands described in Theorem~\ref{thm:conjoint_estimands}. First, we illustrate how anytime-valid CSs and \(p\)-values support principled continuous monitoring of estimates. Second, we show that anytime-valid tests maintain nominal Type I error rates under continuous monitoring, while repeated peeking with fixed-sample tests can severely inflate false-positive risk.
Third, we demonstrate that anytime-valid stopping can yield substantial sample-efficiency gains relative to fixed-sample designs across conjoint settings commonly observed in practice: for moderate effect sizes and sample sizes, the probability of early stopping is often at least 50\% with mean sample savings frequently exceeding 25\%. Fourth, we show that these gains can be especially large when researchers power their studies for effect sizes smaller than those ultimately realized. 
If true effect sizes are even 25\% larger than anticipated, anytime-valid stopping saves more than 20\% of the planned sample size on average.
Finally, we apply the method to a large asylum-seeker conjoint experiment to show how conventional CIs and anytime-valid CSs compare in a real study with many AMCEs.

\subsection{Illustrative example}
\label{sec:illustrative_example}

As a simple illustration, we simulate a forced-choice conjoint experiment between pairs of hypothetical political candidates. Each candidate profile has two attributes:
\begin{itemize}
    \item Candidate region: North (baseline), South, East, West
    \item Candidate political party: Right (baseline), Left
\end{itemize}
Profile attributes are independently and uniformly randomized. We simulate \(G=2{,}500\) respondents, each of whom completes two tasks containing two profiles. This produces 5,000 forced-choice tasks and 10,000 profile-level regression observations. Respondent profile choices are generated using a nonlinear logistic model that includes interactions between party and region and respondent-level preference heterogeneity shared across tasks. The resulting AMCEs are:
\[
\tau_{\text{South}}=-0.0103,\qquad
\tau_{\text{East}}=-0.0768,\qquad
\tau_{\text{West}}=0.0507,\qquad
\tau_{\text{Left}}=0.1001.
\]
Full simulation details are in Appendix~\ref{appendix:sim_details}.

As discussed in Section~\ref{sec:population_projection}, under standard conjoint randomization and identification assumptions, the AMCEs can be estimated by regressing the profile-level outcome---an indicator for whether the profile was chosen---on profile attribute dummy variables, with standard errors clustered at the respondent level.
Applying the cluster-robust methods outlined in Section~\ref{sec:cluster_av_linear_models}, we construct corresponding 95\% CSs and anytime-valid \(p\)-values throughout the experiment.

\begin{figure}[!htbp]
    \centering
    \includegraphics[width=0.8\textwidth]{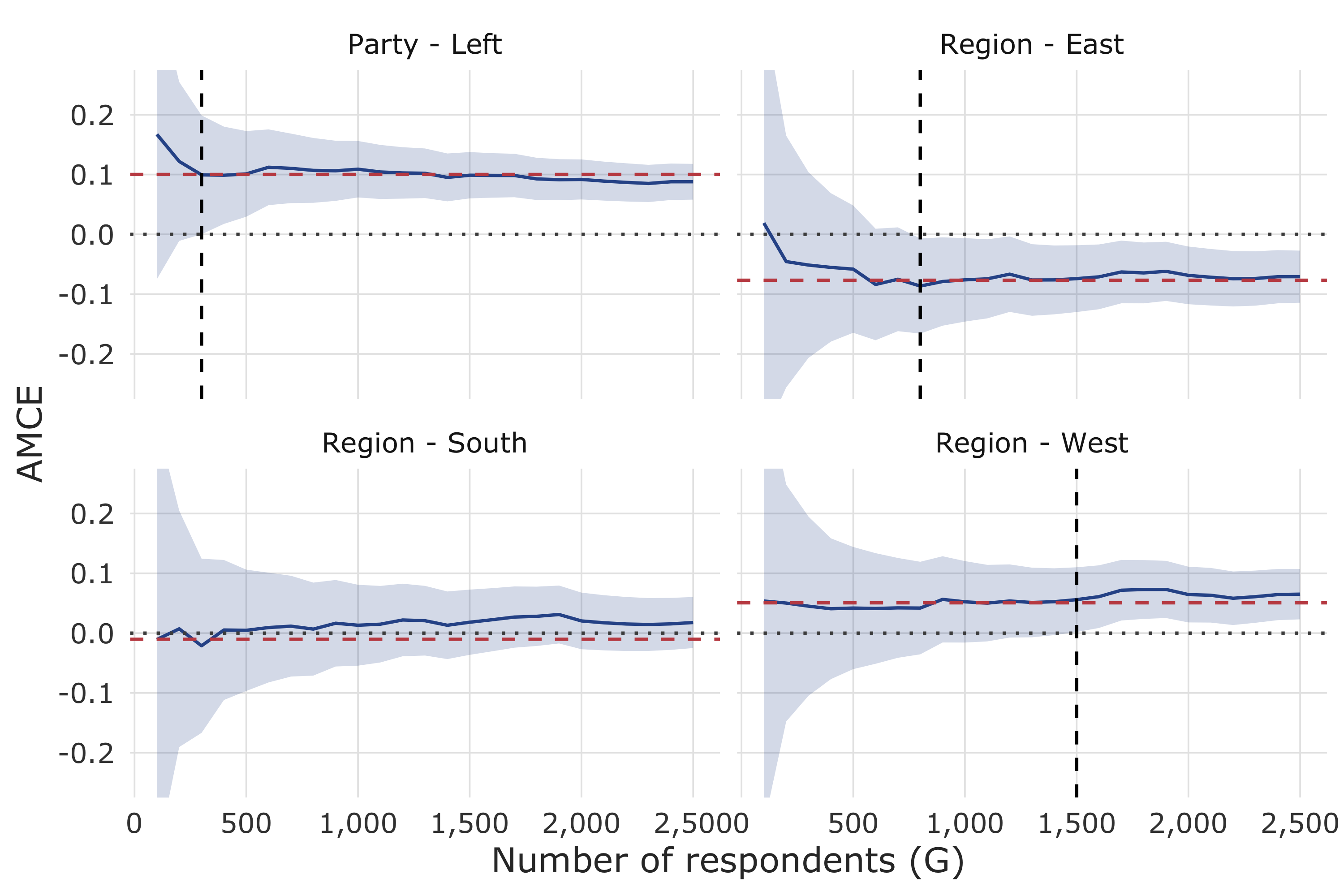}
    \caption{\textbf{AMCE estimates and 95\% confidence sequences.} Red dashed lines denote the true AMCEs under the nonlinear forced-choice DGP, and vertical black dashed lines denote the first respondent sample size \(G\) at which the corresponding confidence sequence excludes 0 (equivalently, when the $p$-value $\leq 0.05$).}
    \label{fig:amce_example}
\end{figure}

Figure~\ref{fig:amce_example} provides a simple illustration of this continuous monitoring process. The estimated AMCEs stabilize around their true values as the number of respondents (\(G\)) grows, while the corresponding CSs shrink over time. Larger effects, such as \(\tau_{\text{Left}}=0.1001\) and \(\tau_{\text{East}}=-0.0768\), become distinguishable from \(0\) relatively early, whereas smaller effects require substantially more respondents and, as in the case of \(\tau_{\text{South}}=-0.0103\), may remain indistinguishable throughout the data collection.

\subsection{Naive peeking inflates false positives}
\label{sec:false_positive_inflation}

Next, we compare anytime-valid and conventional fixed-sample inference under continuous monitoring. We follow the same forced-choice simulation structure as in Section~\ref{sec:illustrative_example}, but impose a null data-generating process in which each profile is chosen with probability \(1/2\) regardless of its attributes, so every AMCE is exactly zero. We consider two null hypotheses. The first tests a single region AMCE, \(H_0:\tau_{\text{South}}=0\). The second jointly tests all three region AMCEs:
\[
H_0:
\tau_{\text{South}}
=
\tau_{\text{East}}
=
\tau_{\text{West}}
=
0.
\]

For each null hypothesis, we calculate both a conventional cluster-robust \(p\)-value and its anytime-valid counterpart from Theorem~\ref{thm:conjoint_estimands} every 10 respondents, beginning at \(G=100\) and continuing through \(G=2{,}500\). Each procedure rejects the null whenever its \(p\)-value \(\leq \alpha=0.05\).
Across 1,000 repeated simulations, we calculate the cumulative probability that each procedure incorrectly rejects the true null at any point during data collection. Additional implementation details are provided in Appendix~\ref{appendix:false_positive_sim_details}.

\begin{figure}[!htbp]
    \centering
    \includegraphics[width=0.9\textwidth]{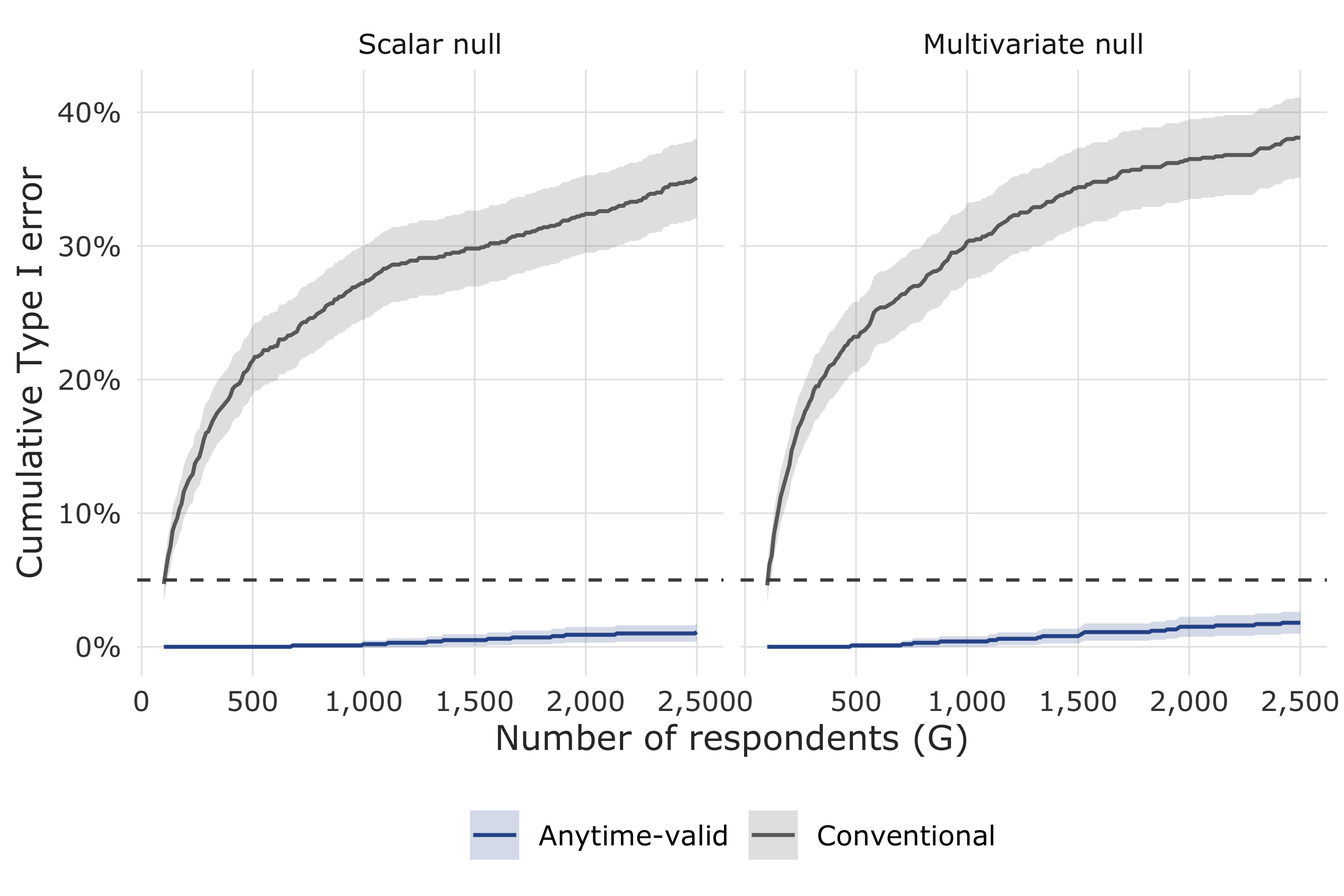}
    \caption{\textbf{Cumulative Type I error rates under sequential monitoring.} The left panel tests a single region AMCE; the right panel jointly tests whether all three non-baseline region AMCEs are zero. The x-axis shows the cumulative number of respondents \(G\), and the y-axis shows the probability of incorrectly rejecting the null at any point up to that value of \(G\). The horizontal dashed line denotes the nominal Type I error level of \(\alpha=0.05\).}
    \label{fig:type1_error}
\end{figure}

Figure~\ref{fig:type1_error} shows that naive continuous monitoring with conventional fixed-sample \(p\)-values severely inflates the probability of a false positive. By \(G=2{,}500\), the conventional procedure has incorrectly rejected the null in approximately 35\% of simulations for the scalar test and 38\% for the joint test. In contrast, the corresponding false-positive rates for the anytime-valid procedure remain below the nominal 5\% level.\footnote{Appendix~\ref{appendix:coverage} reports empirical coverage of the 95\% CSs under the simulation design in Section~\ref{sec:illustrative_example}.}
Figure~\ref{fig:type1_error} reports finite-sample behavior over a limited monitoring horizon. Appendix~\ref{appendix:false_positive_sim_details} extends this empirical exercise through \(G=10^8\).

\subsection{Sample-efficiency gains under plausible conjoint parameters}
\label{sec:sample_efficiency}

\begin{figure}[!htbp]
    \centering
    \makebox[\textwidth][c]{%
        \includegraphics[width=1.2\textwidth]{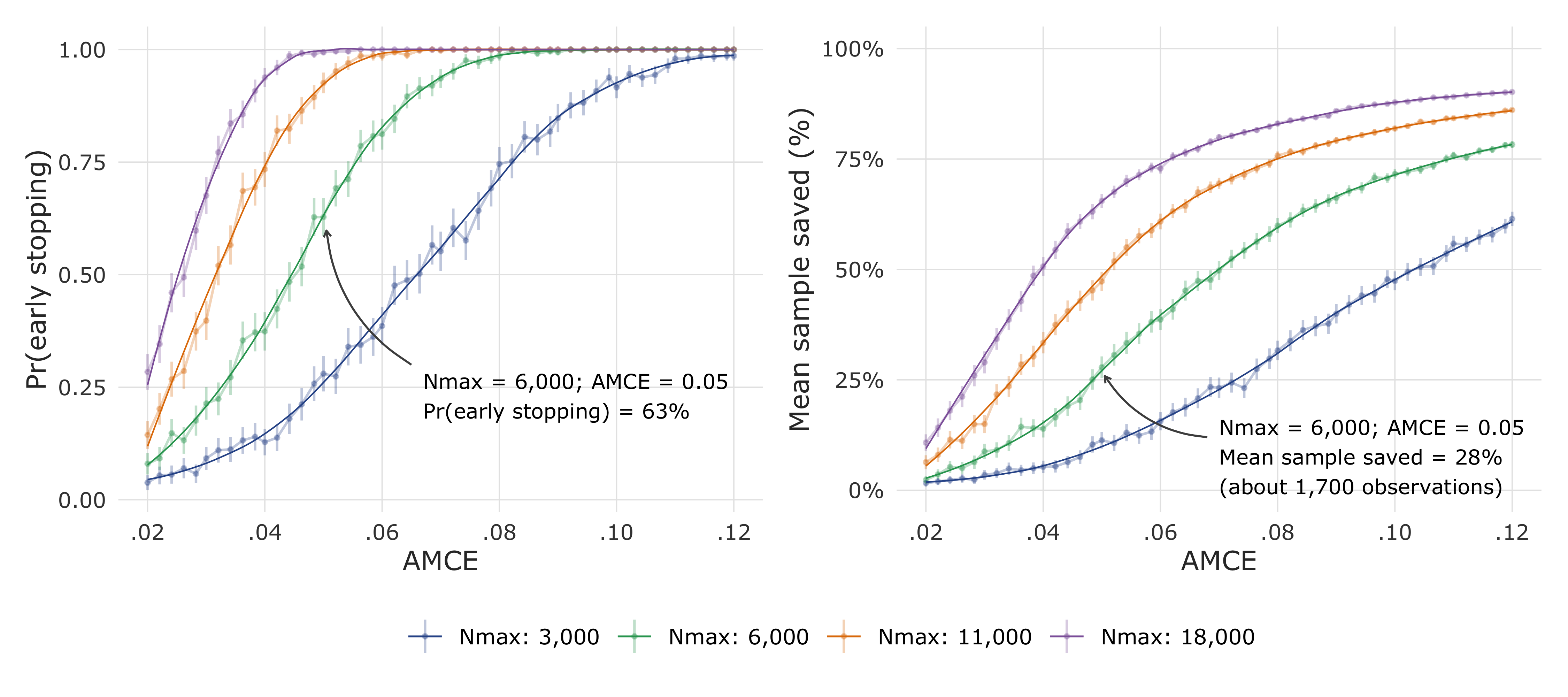}%
    }
    \caption{\textbf{Sample-efficiency gains under anytime-valid stopping.} The left panel shows the probability of early stopping, and the right panel shows the corresponding mean sample savings with 95\% confidence intervals. The x-axis shows the true AMCE effect size for a conjoint with \(n_{\text{levels}}=6\); colors denote maximum effective sample sizes (\(N_{\max}\)).}
    \label{fig:early_stopping}
\end{figure}

Next, we quantify the potential sample-efficiency gains of anytime-valid stopping across a range of conjoint settings commonly observed in practice. Because many conjoint studies do not report having conducted formal power analyses \citep{schuessler_power_2020}, a practically important question is how much sample anytime-valid stopping might save in the kinds of conjoints researchers actually field. To answer this question, we construct a simulation grid of conjoint parameters based on the extensive conjoint literature review of \citet{stefanelli_subjects_2020}.
We consider AMCE values ranging from 0.02 to 0.12, corresponding approximately to the interquartile range of effect sizes reported in that review; maximum effective sample sizes \(N_{\max}\in\{3{,}000,\,6{,}000,\,11{,}000,\,18{,}000\}\) (effective sample size \(=\) number of respondents (G) \(\times\) number of tasks) where \(\left[3{,}000,\,11{,}000\right]\) is the reported interquartile range; and \(n_{\text{levels}}\), the number of levels in the largest attribute, in \(\{4,6,9\}\), where 6 is the reported median and \([4,9]\) is the reported interquartile range.
For each combination of \(\text{AMCE} \times N_{\max} \times n_{\text{levels}}\), we run 500 repeated simulations. Figure~\ref{fig:early_stopping} displays the results for the median case of \(n_{\text{levels}}=6\); the full set of results for \(n_{\text{levels}} \in \{4,6,9\}\) is reported in Appendix~\ref{appendix:sample_efficiency_full}.

We employ a significance-based stopping rule with level \(\alpha = 0.05\): data collection stops when the anytime-valid \(p\)-value for the focal AMCE is \(\leq \alpha\). We summarize the resulting sample-efficiency gains using two quantities: the probability of early stopping, defined as the probability that data collection stops before the planned sample (\(N_{\max}\)) is exhausted, and mean sample savings, defined as the average proportion of the planned sample saved through early stopping.

Figure~\ref{fig:early_stopping} shows that anytime-valid stopping provides substantial gains in both dimensions. The left panel shows that the probability of early stopping increases rapidly as the true effect size grows, while the right panel shows that mean sample savings follow a similar pattern. Even for moderate effect sizes and sample sizes, the probability of early stopping is often at least 50\%, with corresponding mean sample savings that frequently exceed 25\% and can exceed 50\% for somewhat larger effects.

\subsection{Sample-efficiency gains when the true effect exceeds the target}
\label{sec:overshoot}

When using power calculations to choose a sample size, researchers may take a cautious approach by powering for a minimum detectable effect that is a conservative target for the study at hand. If the true effect is larger than this powered-for target, the resulting fixed-sample design may collect substantially more data than is actually needed to achieve the inferential goal. In such cases, anytime-valid stopping can recover a meaningful share of this sample overshoot.

\begin{figure}[!htbp]
    \centering
    \includegraphics[width=0.85\textwidth]{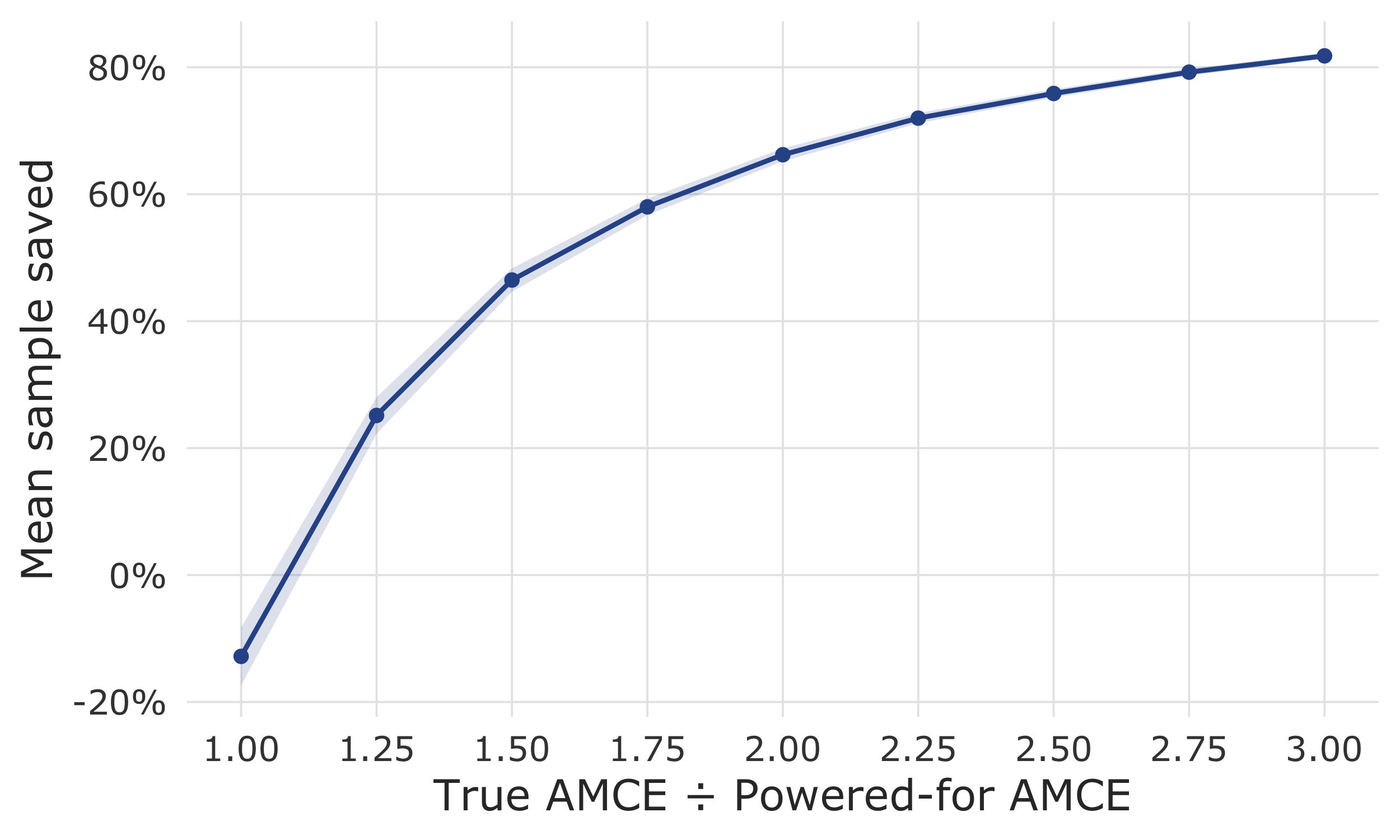}
    \caption{\textbf{Sample savings under power-analysis overshoot.} The x-axis shows the ratio of the true AMCE to the powered-for AMCE, and the y-axis shows the mean sample savings (with 95\% confidence intervals) from anytime-valid stopping relative to the corresponding fixed-sample design.}
    \label{fig:relative_efficiency}
\end{figure}

To quantify the sample-efficiency gains from anytime-valid stopping, we use the power analysis formulae of \citet{schuessler_power_2020} to calculate the effective sample size required to detect a target AMCE with 80\% power in a conjoint with \(n_{\text{levels}}=6\).
We then simulate settings in which the true AMCE exceeds the powered-for target by a ratio ranging from 1 to 3 and apply the same significance-based stopping rule as in Section~\ref{sec:sample_efficiency}.
Figure~\ref{fig:relative_efficiency} reports the resulting mean sample savings from anytime-valid stopping relative to the corresponding fixed-sample design across 1,000 repeated simulations. Full details are given in Appendix~\ref{appendix:overshoot_details}.

Even when the true effect exceeds the powered-for target by only 25\%, anytime-valid stopping recovers more than 20\% of the planned sample size on average. When the true effect is two to three times larger than anticipated, average savings approach 65--80\%. Concretely, suppose a researcher powers their study to detect an AMCE of \(0.01\). If the true AMCE is \(0.02\), they can expect to recover, on average, close to 65\% of the planned sample size.

\subsection{Asylum-seeker conjoint case study}
\label{sec:asylum_case_study}

Finally, we apply our method to the asylum-seeker conjoint experiment conducted by \citet{bansak_how_2016} across 15 European countries. Respondents completed five forced-choice tasks in which they chose between pairs of hypothetical asylum seekers. Profiles varied along nine randomized attributes, including country of origin, age, religion, language skills, vulnerability, and reason for migrating. Using the public replication data released by \citet{bansak_replication_2023}, our analysis includes \(G=18{,}021\) respondents.

We estimate the standard AMCE regression described in Section~\ref{sec:population_projection}, with standard errors clustered at the respondent level. This produces 27 AMCE estimates corresponding to the non-baseline levels of nine randomized attributes. To illustrate how anytime-valid inference could have been used during data collection, we sequentially refit the AMCE regression after every 100 respondents and update the corresponding CSs and anytime-valid \(p\)-values as the sample grows. Additional details are provided in Appendix~\ref{appendix:asylum_case_details}.

\begin{figure}[!htbp]
    \centering
    \includegraphics[width=0.9\textwidth]{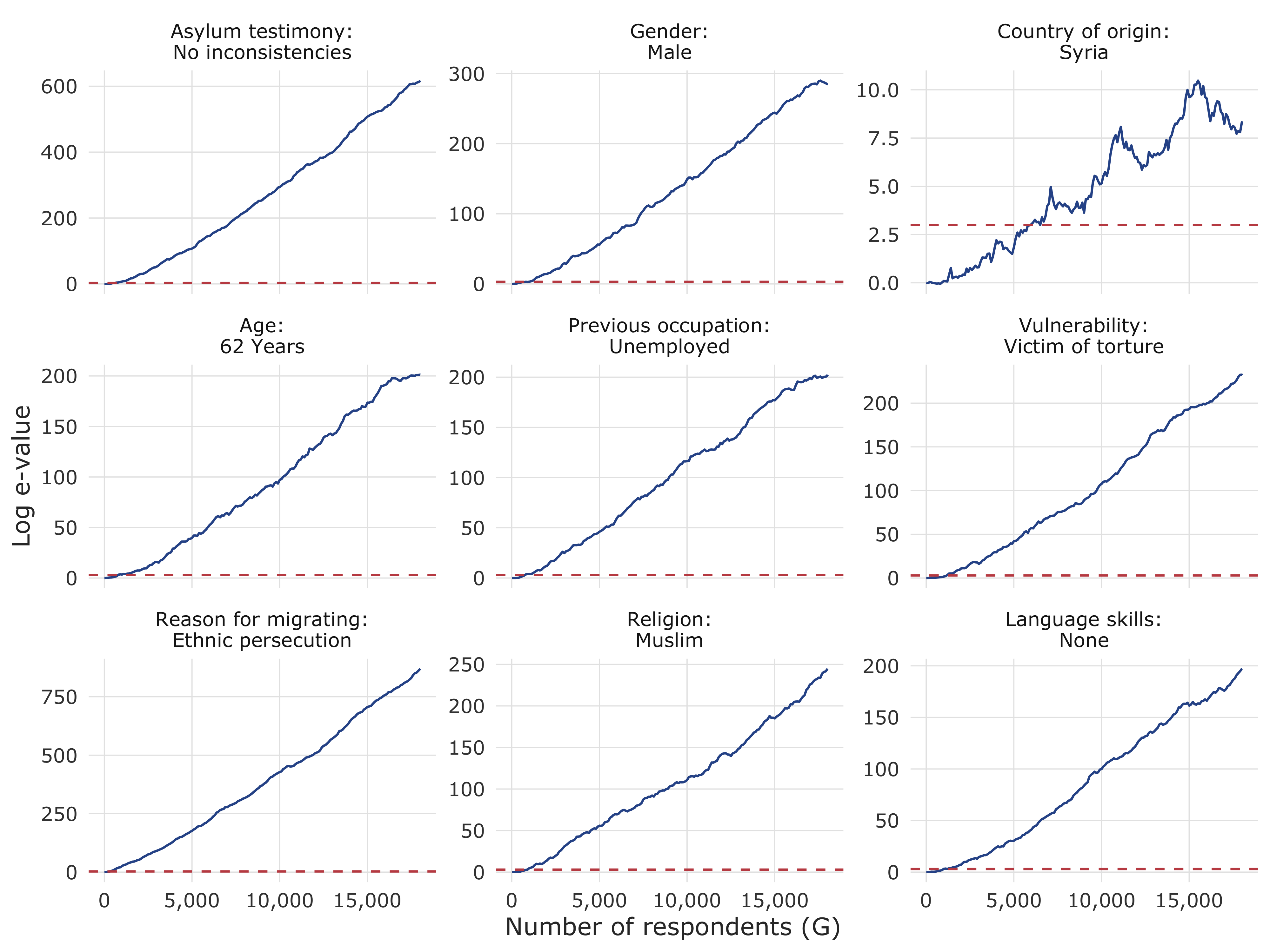}
    \caption{\textbf{Anytime-valid evidence accumulation in the asylum-seeker conjoint case study.} Each panel shows the log \(e\)-value for one AMCE per randomized attribute. The red dashed line denotes the rejection threshold \(\log(1/\alpha)=\log(20)\) at the \(\alpha=0.05\) level.}
    \label{fig:asylum_eprocess_growth}
\end{figure}

Continuous monitoring reveals not only which effects are distinguishable from zero, but also which conclusions emerge quickly and which require substantially more data. Figure~\ref{fig:asylum_eprocess_growth} illustrates this point for a representative subset of the AMCEs, showing that evidence accumulates at substantially different rates across estimands. Each panel plots the log \(e\)-value for an AMCE, representing the strength of evidence against the null hypothesis that the AMCE equals zero. The horizontal red dashed line marks the nominal rejection threshold \(\log(1/\alpha)=\log(20)\) for \(\alpha=0.05\).
The reason-for-migrating AMCE crosses the rejection threshold after \(G \approx 400\) respondents, whereas the Syria country-of-origin AMCE first crosses after \(G \approx 5{,}800\).

\begin{figure}[!htbp]
    \centering
    \includegraphics[width=0.85\textwidth]{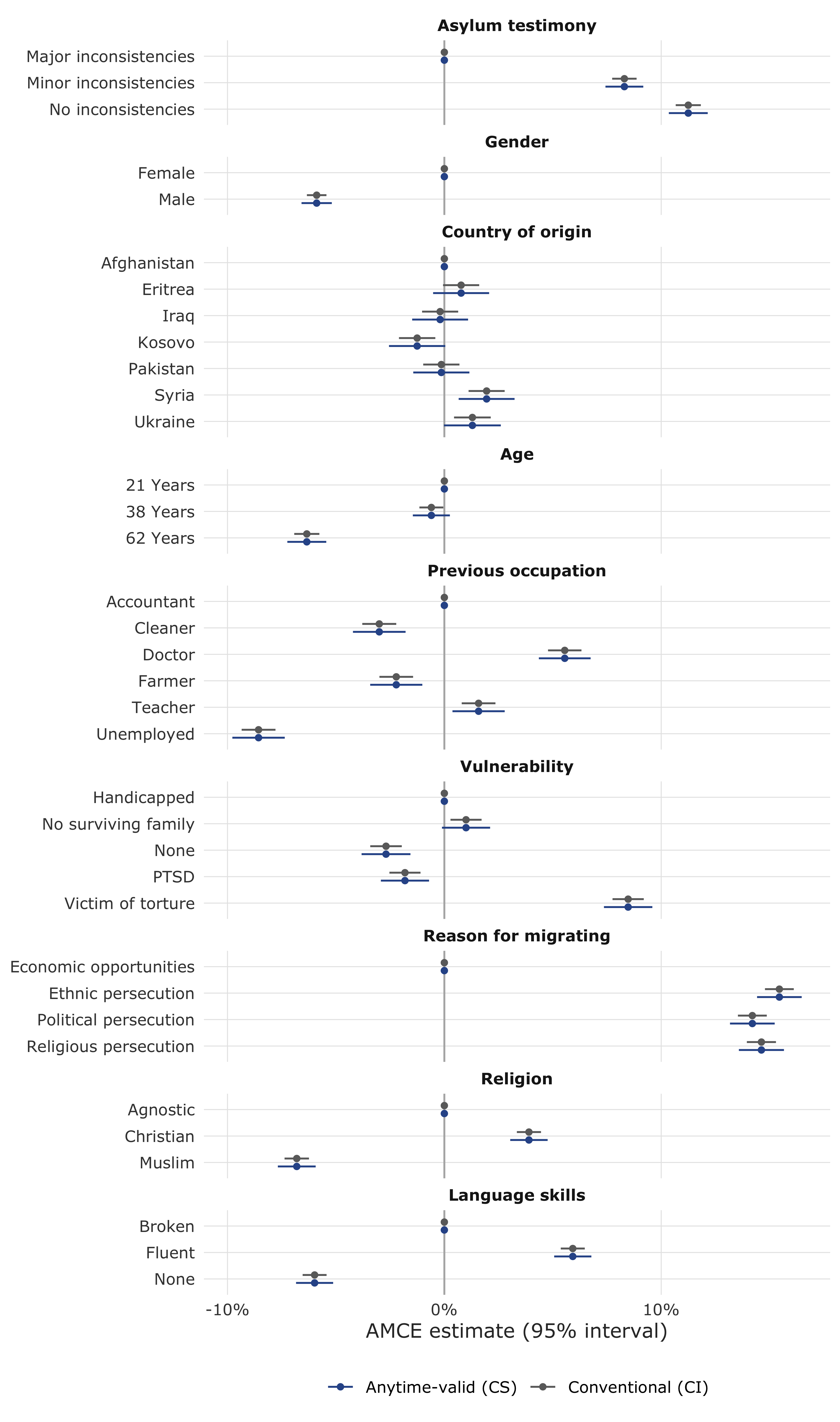}
    \caption{\textbf{Conventional and anytime-valid inference for asylum-seeker AMCEs.} Final AMCE estimates and corresponding conventional 95\% confidence intervals and 95\% anytime-valid confidence sequences. Both use the same OLS estimates and respondent-clustered variance estimator.}
    \label{fig:asylum_case_study}
\end{figure}

Figure~\ref{fig:asylum_case_study} compares the final conventional 95\% CIs and anytime-valid 95\% CSs for all 27 AMCEs.
The CSs are wider because they remain valid under continuous monitoring. At the final sample size, the conventional CIs exclude zero for 24 of the 27 AMCEs, while the anytime-valid CSs exclude zero for 20. The strongest effects involving reason for migrating, testimony consistency, vulnerability, religion, and language skills are clearly distinguishable under both conventional and anytime-valid inference. Thus, the anytime-valid procedure preserves the primary substantive conclusions while differing from conventional inference only for estimates supported by marginal evidence.

\section{Discussion}
\label{sec:discussion}

This paper develops a general framework for anytime-valid inference in conjoint experiments. The primary methodological contribution is to equip a broad class of conjoint estimands with asymptotic \(e\)-processes, CSs, and anytime-valid \(p\)-values.
Many common conjoint estimands are nonparametrically identified under standard assumptions and can be represented as linear contrasts of population least-squares regression coefficients. We formalize the conditions under which the anytime-valid methods of \citet{lindon_clustered_2026} can be applied to this class of estimands.
Finally, we discuss how anytime-valid inference supports data-driven stopping rules, such as rules based on statistical significance and estimation precision, and demonstrate empirically that such rules can yield substantial sample-efficiency gains relative to comparable fixed-sample designs.

In practice, this framework provides a principled alternative to committing in advance to a fixed sample size and refraining from inspecting the data until collection ends. Instead, researchers can monitor accumulating evidence continuously and stop once a pre-specified inferential goal has been reached, without sacrificing valid uncertainty quantification. This flexibility is especially valuable when focal effect sizes are uncertain, when many estimands are of simultaneous interest, or when the operational costs of under- or over-collecting data are substantial.

This work also suggests several directions for future research. One natural direction is to combine anytime-valid inference with recent work on adaptivity in conjoint experiments \citep{offerwestort_adaptive_2021, gosciak_adaptive_2026}. These papers study adaptive treatment assignment to improve experimental efficiency, but do not provide data-driven stopping rules of the kind considered here. Extending anytime-valid inference to such adaptive conjoint designs would complement these methods, allowing researchers to not only adapt how data are collected, but also to determine when enough data have been collected.

A second promising direction concerns adaptivity in the conjoint design itself. Many conjoint estimands, such as AMCEs, are defined relative to an underlying profile distribution \citep{hainmueller_causal_2014, de_la_cuesta_improving_2022}. In some applications, that target distribution may itself be evolving or of substantive interest. This raises the possibility of conjoint designs that adapt elements of the profile distribution over time while using anytime-valid methods to continuously track the estimands induced by those changes. More broadly, anytime-valid inference may serve as a useful building block for future work on adaptive conjoint methodology.

\section*{Funding}

Research reported in this article was supported by the National Science Foundation Graduate Research Fellowship under Grant No. 2139899. Any opinions, findings, and conclusions or recommendations expressed in this material are those of the authors and do not necessarily reflect the views of the National Science Foundation.

\section*{Acknowledgments}

We thank \href{https://www.ianlundberg.org/}{Ian Lundberg}, \href{https://www.skeigold.com/}{Sammy Gold}, and members of the \href{https://koenecke.infosci.cornell.edu/\#lab}{KLEAR lab} for helpful feedback on this project. The authors made use of multiple versions of ChatGPT and Claude between September 2025 and September 2026 to assist with drafting and revising portions of the prose and appendices. All generated text was extensively reviewed and edited by the authors.

\section*{Data availability statement}

A replication package is available at \href{https://github.com/dmolitor/anytime-valid-conjoint}{\textcolor{blue}{github.com/dmolitor/anytime-valid-conjoint}}. The anytime-valid methods presented in this paper are implemented in the open-source R package, \href{https://github.com/michaellindon/avlm}{\textcolor{blue}{avlm}}.

\newpage

\bibliography{references}

@article{hainmueller_causal_2014,
	title = {Causal {Inference} in {Conjoint} {Analysis}: {Understanding} {Multidimensional} {Choices} via {Stated} {Preference} {Experiments}},
	volume = {22},
	copyright = {https://www.cambridge.org/core/terms},
	issn = {1047-1987, 1476-4989},
	shorttitle = {Causal {Inference} in {Conjoint} {Analysis}},
	url = {https://www.cambridge.org/core/product/identifier/S1047198700013589/type/journal_article},
	doi = {10.1093/pan/mpt024},
	language = {en},
	number = {1},
	urldate = {2025-04-08},
	journal = {Political Analysis},
	author = {Hainmueller, Jens and Hopkins, Daniel J. and Yamamoto, Teppei},
	year = {2014},
	pages = {1--30},
}

@misc{stefanelli_subjects_2020,
	title = {Subjects, {Trials}, and {Levels}: {Statistical} {Power} in {Conjoint} {Experiments}},
	shorttitle = {Subjects, {Trials}, and {Levels}},
	url = {https://osf.io/spkcy},
	doi = {10.31235/osf.io/spkcy},
	urldate = {2025-04-09},
	publisher = {SocArXiv},
	author = {Stefanelli, Alberto and Lukac, Martin},
	month = nov,
	year = {2020},
    howpublished = {SocArXiv. \url{https://doi.org/10.31235/osf.io/spkcy}},
}

@misc{schuessler_power_2020,
	title = {Power {Analysis} for {Conjoint} {Experiments}},
	copyright = {https://creativecommons.org/licenses/by/4.0/legalcode},
	url = {https://osf.io/9yuhp},
	doi = {10.31235/osf.io/9yuhp},
	urldate = {2025-04-22},
	publisher = {SocArXiv},
	author = {Schuessler, Julian and Freitag, Markus},
	month = dec,
	year = {2020},
    howpublished = {SocArXiv. \url{https://doi.org/10.31235/osf.io/9yuhp}},
}

@article{leeper_measuring_2020,
	title = {Measuring {Subgroup} {Preferences} in {Conjoint} {Experiments}},
	volume = {28},
	copyright = {https://www.cambridge.org/core/terms},
	issn = {1047-1987, 1476-4989},
	url = {https://www.cambridge.org/core/product/identifier/S1047198719000305/type/journal_article},
	doi = {10.1017/pan.2019.30},
	language = {en},
	number = {2},
	urldate = {2025-04-22},
	journal = {Political Analysis},
	author = {Leeper, Thomas J. and Hobolt, Sara B. and Tilley, James},
	month = apr,
	year = {2020},
	pages = {207--221},
}

@article{de_la_cuesta_improving_2022,
	title = {Improving the {External} {Validity} of {Conjoint} {Analysis}: {The} {Essential} {Role} of {Profile} {Distribution}},
	volume = {30},
	copyright = {https://www.cambridge.org/core/terms},
	issn = {1047-1987, 1476-4989},
	shorttitle = {Improving the {External} {Validity} of {Conjoint} {Analysis}},
	url = {https://www.cambridge.org/core/product/identifier/S1047198720000406/type/journal_article},
	doi = {10.1017/pan.2020.40},
	language = {en},
	number = {1},
	urldate = {2025-09-17},
	journal = {Political Analysis},
	author = {De La Cuesta, Brandon and Egami, Naoki and Imai, Kosuke},
	month = jan,
	year = {2022},
	pages = {19--45},
}

@article{egami_causal_2019,
	title = {Causal {Interaction} in {Factorial} {Experiments}: {Application} to {Conjoint} {Analysis}},
	volume = {114},
	issn = {0162-1459, 1537-274X},
	shorttitle = {Causal {Interaction} in {Factorial} {Experiments}},
	url = {https://www.tandfonline.com/doi/full/10.1080/01621459.2018.1476246},
	doi = {10.1080/01621459.2018.1476246},
	language = {en},
	number = {526},
	urldate = {2025-09-17},
	journal = {Journal of the American Statistical Association},
	author = {Egami, Naoki and Imai, Kosuke},
	month = apr,
	year = {2019},
	pages = {529--540},
}

@article{gelman_beyond_2014,
	title = {Beyond {Power} {Calculations}: {Assessing} {Type} {S} ({Sign}) and {Type} {M} ({Magnitude}) {Errors}},
	volume = {9},
	issn = {1745-6916, 1745-6924},
	shorttitle = {Beyond {Power} {Calculations}},
	url = {https://journals.sagepub.com/doi/10.1177/1745691614551642},
	doi = {10.1177/1745691614551642},
	language = {en},
	number = {6},
	urldate = {2026-03-31},
	journal = {Perspectives on Psychological Science},
	author = {Gelman, Andrew and Carlin, John},
	month = nov,
	year = {2014},
	pages = {641--651},
}

@article{liu_multiple_2023,
	title = {Multiple {Hypothesis} {Testing} in {Conjoint} {Analysis}},
	volume = {31},
	copyright = {https://creativecommons.org/licenses/by/4.0},
	issn = {1047-1987, 1476-4989},
	url = {https://www.cambridge.org/core/product/identifier/S1047198722000304/type/journal_article},
	doi = {10.1017/pan.2022.30},
	language = {en},
	number = {3},
	urldate = {2026-03-31},
	journal = {Political Analysis},
	author = {Liu, Guoer and Shiraito, Yuki},
	month = jul,
	year = {2023},
	pages = {380--395},
}

@article{howard_time-uniform_2021,
	title = {Time-uniform, nonparametric, nonasymptotic confidence sequences},
	volume = {49},
	issn = {0090-5364},
	url = {https://projecteuclid.org/journals/annals-of-statistics/volume-49/issue-2/Time-uniform-nonparametric-nonasymptotic-confidence-sequences/10.1214/20-AOS1991.full},
	doi = {10.1214/20-AOS1991},
	number = {2},
	urldate = {2026-03-31},
	journal = {The Annals of Statistics},
	author = {Howard, Steven R. and Ramdas, Aaditya and McAuliffe, Jon and Sekhon, Jasjeet},
	month = apr,
	year = {2021},
}

@article{simmons_false-positive_2011,
	title = {False-{Positive} {Psychology}: {Undisclosed} {Flexibility} in {Data} {Collection} and {Analysis} {Allows} {Presenting} {Anything} as {Significant}},
	volume = {22},
	issn = {0956-7976, 1467-9280},
	shorttitle = {False-{Positive} {Psychology}},
	url = {https://journals.sagepub.com/doi/10.1177/0956797611417632},
	doi = {10.1177/0956797611417632},
	language = {en},
	number = {11},
	urldate = {2026-03-31},
	journal = {Psychological Science},
	author = {Simmons, Joseph P. and Nelson, Leif D. and Simonsohn, Uri},
	month = nov,
	year = {2011},
	pages = {1359--1366},
}

@article{ramdas_hypothesis_2025,
	title = {Hypothesis testing with e-values},
	volume = {1},
	issn = {2978-4212, 2978-4220},
	url = {http://arxiv.org/abs/2410.23614},
	doi = {10.1561/3600000002},
	number = {1-2},
	urldate = {2026-04-01},
	journal = {Foundations and Trends® in Statistics},
	author = {Ramdas, Aaditya and Wang, Ruodu},
	month = jul,
	year = {2025},
	note = {arXiv:2410.23614 [math]},
	pages = {1--390},
}

@misc{ramdas_game-theoretic_2023,
	title = {Game-theoretic statistics and safe anytime-valid inference},
	url = {http://arxiv.org/abs/2210.01948},
	doi = {10.48550/arXiv.2210.01948},
	urldate = {2026-04-01},
	publisher = {arXiv},
	author = {Ramdas, Aaditya and Grünwald, Peter and Vovk, Vladimir and Shafer, Glenn},
	month = jun,
	year = {2023},
	note = {arXiv:2210.01948 [math]},
}

@article{xu_post-selection_2024,
	title = {Post-selection inference for e-value based confidence intervals},
	volume = {18},
	issn = {1935-7524},
	url = {https://projecteuclid.org/journals/electronic-journal-of-statistics/volume-18/issue-1/Post-selection-inference-for-e-value-based-confidence-intervals/10.1214/24-EJS2253.full},
	doi = {10.1214/24-EJS2253},
	number = {1},
	urldate = {2026-04-13},
	journal = {Electronic Journal of Statistics},
	author = {Xu, Ziyu and Wang, Ruodu and Ramdas, Aaditya},
	month = jan,
	year = {2024},
}

@article{benjamini_false_2005,
	title = {False {Discovery} {Rate}–{Adjusted} {Multiple} {Confidence} {Intervals} for {Selected} {Parameters}},
	volume = {100},
	issn = {0162-1459, 1537-274X},
	url = {http://www.tandfonline.com/doi/abs/10.1198/016214504000001907},
	doi = {10.1198/016214504000001907},
	language = {en},
	number = {469},
	urldate = {2026-04-13},
	journal = {Journal of the American Statistical Association},
	author = {Benjamini, Yoav and Yekutieli, Daniel},
	month = mar,
	year = {2005},
	pages = {71--81},
}

@article{dunn_multiple_1961,
	title = {Multiple {Comparisons} among {Means}},
	volume = {56},
	issn = {0162-1459, 1537-274X},
	url = {http://www.tandfonline.com/doi/abs/10.1080/01621459.1961.10482090},
	doi = {10.1080/01621459.1961.10482090},
	language = {en},
	number = {293},
	urldate = {2026-04-13},
	journal = {Journal of the American Statistical Association},
	author = {Dunn, Olive Jean},
	month = mar,
	year = {1961},
	pages = {52--64},
}

@article{benjamini_controlling_1995,
	title = {Controlling the {False} {Discovery} {Rate}: {A} {Practical} and {Powerful} {Approach} to {Multiple} {Testing}},
	volume = {57},
	copyright = {https://academic.oup.com/journals/pages/open\_access/funder\_policies/chorus/standard\_publication\_model},
	issn = {1369-7412, 1467-9868},
	shorttitle = {Controlling the {False} {Discovery} {Rate}},
	url = {https://academic.oup.com/jrsssb/article/57/1/289/7035855},
	doi = {10.1111/j.2517-6161.1995.tb02031.x},
	language = {en},
	number = {1},
	urldate = {2026-04-13},
	journal = {Journal of the Royal Statistical Society Series B: Statistical Methodology},
	author = {Benjamini, Yoav and Hochberg, Yosef},
	month = jan,
	year = {1995},
	pages = {289--300},
}

@article{johari_always_2022,
	title = {Always {Valid} {Inference}: {Continuous} {Monitoring} of {A}/{B} {Tests}},
	volume = {70},
	issn = {0030-364X, 1526-5463},
	shorttitle = {Always {Valid} {Inference}},
	url = {https://pubsonline.informs.org/doi/10.1287/opre.2021.2135},
	doi = {10.1287/opre.2021.2135},
	language = {en},
	number = {3},
	urldate = {2026-04-13},
	journal = {Operations Research},
	author = {Johari, Ramesh and Koomen, Pete and Pekelis, Leonid and Walsh, David},
	month = may,
	year = {2022},
	pages = {1806--1821},
}

@article{gosciak_adaptive_2026,
	title = {Adaptive {Randomization} in {Conjoint} {Survey} {Experiments}},
	volume = {34},
	copyright = {https://creativecommons.org/licenses/by-nc-nd/4.0},
	issn = {1047-1987, 1476-4989},
	url = {https://www.cambridge.org/core/product/identifier/S1047198726100400/type/journal_article},
	doi = {10.1017/pan.2026.10040},
	language = {en},
	number = {4},
	urldate = {2026-09-29},
	journal = {Political Analysis},
	author = {Gosciak, Jennah and Molitor, Daniel and Lundberg, Ian},
	month = oct,
	year = {2026},
	pages = {541--568},
}

@article{offerwestort_adaptive_2021,
	title = {Adaptive {Experimental} {Design}: {Prospects} and {Applications} in {Political} {Science}},
	volume = {65},
	issn = {0092-5853, 1540-5907},
	shorttitle = {Adaptive {Experimental} {Design}},
	url = {https://onlinelibrary.wiley.com/doi/10.1111/ajps.12597},
	doi = {10.1111/ajps.12597},
	language = {en},
	number = {4},
	urldate = {2026-04-13},
	journal = {American Journal of Political Science},
	author = {Offer‐Westort, Molly and Coppock, Alexander and Green, Donald P.},
	month = oct,
	year = {2021},
	pages = {826--844},
}

@misc{liang_experimental_2024,
	title = {An {Experimental} {Design} for {Anytime}-{Valid} {Causal} {Inference} on {Multi}-{Armed} {Bandits}},
	url = {http://arxiv.org/abs/2311.05794},
	doi = {10.48550/arXiv.2311.05794},
	urldate = {2026-04-14},
	publisher = {arXiv},
	author = {Liang, Biyonka and Bojinov, Iavor},
	month = oct,
	year = {2024},
	note = {arXiv:2311.05794 [stat]},
}

@misc{ham_design-based_2023,
	title = {Design-{Based} {Confidence} {Sequences}: {A} {General} {Approach} to {Risk} {Mitigation} in {Online} {Experimentation}},
	shorttitle = {Design-{Based} {Confidence} {Sequences}},
	url = {http://arxiv.org/abs/2210.08639},
	doi = {10.48550/arXiv.2210.08639},
	urldate = {2026-04-14},
	publisher = {arXiv},
	author = {Ham, Dae Woong and Bojinov, Iavor and Lindon, Michael and Tingley, Martin},
	month = may,
	year = {2023},
	note = {arXiv:2210.08639 [stat]},
}

@misc{molitor_anytime-valid_2025,
	title = {Anytime-{Valid} {Inference} in {Adaptive} {Experiments}: {Covariate} {Adjustment} and {Balanced} {Power}},
	shorttitle = {Anytime-{Valid} {Inference} in {Adaptive} {Experiments}},
	url = {http://arxiv.org/abs/2506.20523},
	doi = {10.48550/arXiv.2506.20523},
	urldate = {2026-04-14},
	publisher = {arXiv},
	author = {Molitor, Daniel and Gold, Samantha},
	month = jun,
	year = {2025},
	note = {arXiv:2506.20523 [stat]},
}

@article{cunningham_adaptive_2010,
	title = {Adaptive {Choice}-{Based} {Conjoint} {Analysis}: {A} {New} {Patient}-{Centered} {Approach} to the {Assessment} of {Health} {Service} {Preferences}},
	volume = {3},
	issn = {1178-1653},
	shorttitle = {Adaptive {Choice}-{Based} {Conjoint} {Analysis}},
	url = {http://link.springer.com/10.2165/11537870-000000000-00000},
	doi = {10.2165/11537870-000000000-00000},
	language = {en},
	number = {4},
	urldate = {2026-04-14},
	journal = {The Patient: Patient-Centered Outcomes Research},
	author = {Cunningham, Charles E. and Deal, Ken and Chen, Yvonne},
	month = dec,
	year = {2010},
	pages = {257--273},
}

@article{yu_individually_2011,
	title = {Individually adapted sequential {Bayesian} conjoint-choice designs in the presence of consumer heterogeneity},
	volume = {28},
	copyright = {https://www.elsevier.com/tdm/userlicense/1.0/},
	issn = {01678116},
	url = {https://linkinghub.elsevier.com/retrieve/pii/S0167811611000668},
	doi = {10.1016/j.ijresmar.2011.06.002},
	language = {en},
	number = {4},
	urldate = {2026-04-14},
	journal = {International Journal of Research in Marketing},
	author = {Yu, Jie and Goos, Peter and Vandebroek, Martina},
	month = dec,
	year = {2011},
	pages = {378--388},
}

@article{frey_fixed-width_2010,
	title = {Fixed-{Width} {Sequential} {Confidence} {Intervals} for a {Proportion}},
	volume = {64},
	issn = {0003-1305, 1537-2731},
	url = {http://www.tandfonline.com/doi/abs/10.1198/tast.2010.09140},
	doi = {10.1198/tast.2010.09140},
	language = {en},
	number = {3},
	urldate = {2026-05-16},
	journal = {The American Statistician},
	author = {Frey, Jesse},
	month = aug,
	year = {2010},
	pages = {242--249},
}

@misc{bansak_replication_2023,
	title = {Replication {Data} for: {Europeans}' support for refugees of varying background is stable over time},
	copyright = {, Creative Commons Zero v1.0 Universal},
	shorttitle = {Replication {Data} for},
	url = {https://dataverse.harvard.edu/citation?persistentId=doi:10.7910/DVN/FTL1MM},
	doi = {10.7910/DVN/FTL1MM},
	urldate = {2026-08-26},
	publisher = {Harvard Dataverse},
	author = {Bansak, Kirk and Hainmueller, Jens and Hangartner, Dominik},
	collaborator = {Bansak, Kirk},
	year = {2023},
	note = {Artwork Size: 5975989, 5976121
Pages: 5975989, 5976121},
}

@article{lindon_anytime-valid_2026,
	title = {Anytime-{Valid} {Inference} in {Linear} {Models} with {Applications} to {Regression}-{Adjusted} {Causal} {Inference}},
	issn = {0162-1459, 1537-274X},
	url = {https://www.tandfonline.com/doi/full/10.1080/01621459.2026.2692052},
	doi = {10.1080/01621459.2026.2692052},
	language = {en},
	urldate = {2026-09-14},
	journal = {Journal of the American Statistical Association},
	author = {Lindon, Michael and Ham, Dae Woong and Tingley, Martin and Bojinov, Iavor},
	month = aug,
	year = {2026},
	pages = {1--12},
}

@article{bansak_how_2016,
	title = {How economic, humanitarian, and religious concerns shape {European} attitudes toward asylum seekers},
	volume = {354},
	copyright = {http://www.sciencemag.org/about/science-licenses-journal-article-reuse},
	issn = {0036-8075, 1095-9203},
	url = {https://www.science.org/doi/10.1126/science.aag2147},
	doi = {10.1126/science.aag2147},
	language = {en},
	number = {6309},
	urldate = {2026-09-16},
	journal = {Science},
	author = {Bansak, Kirk and Hainmueller, Jens and Hangartner, Dominik},
	month = oct,
	year = {2016},
	pages = {217--222},
}

@article{bester_inference_2011,
	title = {Inference with dependent data using cluster covariance estimators},
	volume = {165},
	copyright = {https://www.elsevier.com/tdm/userlicense/1.0/},
	issn = {03044076},
	url = {https://linkinghub.elsevier.com/retrieve/pii/S0304407611000431},
	doi = {10.1016/j.jeconom.2011.01.007},
	language = {en},
	number = {2},
	urldate = {2026-09-16},
	journal = {Journal of Econometrics},
	author = {Bester, C. Alan and Conley, Timothy G. and Hansen, Christian B.},
	month = dec,
	year = {2011},
	pages = {137--151},
}

@article{colin_cameron_practitioners_2015,
	title = {A {Practitioner}’s {Guide} to {Cluster}-{Robust} {Inference}},
	volume = {50},
	issn = {0022-166X, 1548-8004},
	url = {http://jhr.uwpress.org/lookup/doi/10.3368/jhr.50.2.317},
	doi = {10.3368/jhr.50.2.317},
	language = {en},
	number = {2},
	urldate = {2026-09-16},
	journal = {Journal of Human Resources},
	author = {Colin Cameron, A. and Miller, Douglas L.},
	year = {2015},
	pages = {317--372},
}

@article{mackinnon_cluster-robust_2023,
	title = {Cluster-robust inference: {A} guide to empirical practice},
	volume = {232},
	issn = {03044076},
	shorttitle = {Cluster-robust inference},
	url = {https://linkinghub.elsevier.com/retrieve/pii/S0304407622000781},
	doi = {10.1016/j.jeconom.2022.04.001},
	language = {en},
	number = {2},
	urldate = {2026-09-16},
	journal = {Journal of Econometrics},
	author = {MacKinnon, James G. and Nielsen, Morten {\O}rregaard and Webb, Matthew D.},
	month = feb,
	year = {2023},
	pages = {272--299},
}

@article{pustejovsky_small-sample_2016,
	title = {Small-{Sample} {Methods} for {Cluster}-{Robust} {Variance} {Estimation} and {Hypothesis} {Testing} in {Fixed} {Effects} {Models}},
	volume = {36},
	issn = {0735-0015, 1537-2707},
	url = {https://www.tandfonline.com/doi/full/10.1080/07350015.2016.1247004},
	doi = {10.1080/07350015.2016.1247004},
	language = {en},
	number = {4},
	urldate = {2026-09-16},
	journal = {Journal of Business \& Economic Statistics},
	author = {Pustejovsky, James E. and Tipton, Elizabeth},
	month = oct,
	year = {2016},
	pages = {672--683},
}

@article{wang_false_2022,
	title = {False {Discovery} {Rate} {Control} with {E}-values},
	volume = {84},
	copyright = {https://academic.oup.com/journals/pages/open\_access/funder\_policies/chorus/standard\_publication\_model},
	issn = {1369-7412, 1467-9868},
	url = {https://academic.oup.com/jrsssb/article/84/3/822/7056146},
	doi = {10.1111/rssb.12489},
	language = {en},
	number = {3},
	urldate = {2026-09-16},
	journal = {Journal of the Royal Statistical Society Series B: Statistical Methodology},
	author = {Wang, Ruodu and Ramdas, Aaditya},
	month = jul,
	year = {2022},
	pages = {822--852},
}

@article{waudby-smith_time-uniform_2024,
	title = {Time-uniform central limit theory and asymptotic confidence sequences},
	volume = {52},
	issn = {0090-5364},
	url = {https://projecteuclid.org/journals/annals-of-statistics/volume-52/issue-6/Time-uniform-central-limit-theory-and-asymptotic-confidence-sequences/10.1214/24-AOS2408.full},
	doi = {10.1214/24-AOS2408},
	number = {6},
	urldate = {2026-09-16},
	journal = {The Annals of Statistics},
	author = {Waudby-Smith, Ian and Arbour, David and Sinha, Ritwik and Kennedy, Edward H. and Ramdas, Aaditya},
	month = dec,
	year = {2024},
}

@inproceedings{cook_semiparametric_2024,
	title = {Semiparametric {Efficient} {Inference} in {Adaptive} {Experiments}},
	issn = {2640-3498},
	url = {https://proceedings.mlr.press/v236/cook24a.html},
	language = {en},
	urldate = {2026-10-01},
	booktitle = {Proceedings of the {Third} {Conference} on {Causal} {Learning} and {Reasoning}},
	publisher = {PMLR},
	author = {Cook, Thomas and Mishler, Alan and Ramdas, Aaditya},
	month = mar,
	year = {2024},
	pages = {1033--1064},
}

@misc{lindon_clustered_2026,
  title={Anytime-Valid Inference in Linear Models with Clustered Outcomes: Exact Multivariate Tests and Cluster-Robust Confidence Sequences},
  author={Lindon, Michael},
  year={2026},
  eprint={2610.25613},
  archivePrefix={arXiv},
  primaryClass={stat.ME},
  howpublished={arXiv:2610.25613v1 [stat.ME]},
  url={https://arxiv.org/abs/2610.25613}
}

\newpage

\appendix

\section{Proof of Proposition~\ref{prop:amce_projection}}
\label{appendix:amce_projection_proof}

For the reader's edification, we provide a proof of Proposition~\ref{prop:amce_projection}. For the underlying AMCE identification result, see \citet[Proposition~3]{hainmueller_causal_2014}.

\begin{proof}
We first establish the equivalence between the respondent-level projection in Definition~\ref{def:projection} and the profile-level regression used in Proposition~\ref{prop:amce_projection}. Let \(N_i\) denote the number of profile evaluations contributed by respondent \(i\), and let \(x_{ir}\) denote the regressor vector for profile evaluation \(r\). The respondent-level objective is
\[
\mathbb E_P\!\left[\|Y_i-X_i\beta\|_2^2\right]
=
\mathbb E_P\!\left[
\sum_{r=1}^{N_i}
(Y_{ir}-x_{ir}'\beta)^2
\right].
\]
Under the conditions of Section~\ref{sec:amce}, each profile-level pair \((Y_{ir},x_{ir})\) has the same marginal distribution. Because \(N_i\) is independent of the profile assignments and potential outcomes,
\[
\mathbb E_P\!\left[\|Y_i-X_i\beta\|_2^2\right]
=
\mathbb E[N_i]\,
\mathbb E_P\!\left[(Y-x'\beta)^2\right].
\]
The respondent-level and profile-level objectives therefore differ only by the positive constant \(\mathbb E[N_i]\) and have the same minimizer. This result does not require outcomes contributed by the same respondent to be independent; within-respondent dependence affects the covariance of the estimator, not the population coefficient vector. The required moments exist automatically for the bounded outcomes and indicator regressors considered here.

We may therefore drop the subscripts \(ikm\) and write \(Y\) for a generic profile-level outcome and \(A=(A_1,\ldots,A_J)\sim\pi\) for its attribute vector. For each attribute \(j\), collect the non-reference indicators in the vector
\(
W_j
=
\left(
\mathbf{1}\{A_j=\ell\}
\right)_{\ell\in\mathcal{L}_j\setminus\{\ell_{0}\}},
\)
and let \(\widetilde W_j=W_j-\mathbb E(W_j)\). The regressor vector is \((1,W_1,\ldots,W_J)\), and the coefficient vector is \(\beta=(\alpha,\beta_1,\ldots,\beta_J)\), where
\(
\beta_j
=
\left(
\beta_{j_\ell}
\right)_{\ell\in\mathcal{L}_j\setminus\{\ell_{0}\}}.
\)

Two facts follow from complete independent randomization. First, because \(A_1,\ldots,A_J\) are mutually independent and \(W_j\) is a function of \(A_j\) alone, the centered blocks are orthogonal across attributes: \(\mathbb E(\widetilde W_j\widetilde W_{j'}')=0\) for \(j\neq j'\). Second, because every level of attribute \(j\) has positive probability, the matrix
\(
\mathbb E(\widetilde W_j\widetilde W_j')
=\operatorname{Cov}(W_j)
=\operatorname{diag}(p_j)-p_jp_j',
\)
where \(p_j\) collects the probabilities of the non-reference levels of attribute \(j\), is positive definite: the entries of \(p_j\) are positive and, since the reference level also has positive probability, sum to less than one.

Reparametrize the regression function as
\[
\alpha+\sum_{j=1}^J W_j'\beta_j
=
c+\sum_{j=1}^J\widetilde W_j'\beta_j,
\qquad
c=\alpha+\sum_{j=1}^J\mathbb E(W_j)'\beta_j,
\]
which is a bijection between \((\alpha,\beta_1,\ldots,\beta_J)\) and \((c,\beta_1,\ldots,\beta_J)\). Using \(\mathbb E(\widetilde W_j)=0\) and the orthogonality across blocks, the population least-squares objective separates:
\[
\mathbb E\Big[\Big(Y-c-\sum_{j=1}^J\widetilde W_j'\beta_j\Big)^2\Big]
=
\mathbb E\big[(Y-c)^2\big]
+\sum_{j=1}^J\Big\{
\beta_j'\,\mathbb E(\widetilde W_j\widetilde W_j')\,\beta_j
-2\,\beta_j'\,\mathbb E(\widetilde W_jY)
\Big\}.
\]
Each term depends on a separate block of parameters, so the unique minimizer is
\[
c^\star=\mathbb E(Y),
\qquad
\beta_j^\star
=\mathbb E(\widetilde W_j\widetilde W_j')^{-1}\,\mathbb E(\widetilde W_jY)
=\operatorname{Cov}(W_j)^{-1}\operatorname{Cov}(W_j,Y),
\qquad j=1,\ldots,J.
\]
In particular, \(\beta^\star\) is unique, so the positive-definiteness requirement noted after Definition~\ref{def:projection} is satisfied.

The vector \(\beta_j^\star\) is exactly the slope vector of the population least-squares regression of \(Y\) on \((1,W_j)\), that is, on the indicators of attribute \(j\) alone. Because every function of \(A_j\) can be written as \(f(A_j)=f(\ell_{0})+\sum_{\ell\neq\ell_{0}}\{f(\ell)-f(\ell_{0})\}\mathbf{1}\{A_j=\ell\}\), the linear span of \((1,W_j)\) is the space of all functions of \(A_j\), and the population projection of \(Y\) onto that space is the conditional mean \(\mathbb E(Y\mid A_j)\). Hence there is a constant \(c_j\) such that \(c_j+W_j'\beta_j^\star=\mathbb E(Y\mid A_j)\) almost surely. Evaluating at \(A_j=\ell_{0}\), where \(W_j=0\), gives \(c_j=\mathbb E(Y\mid A_j=\ell_{0})\); evaluating at \(A_j=\ell\) for \(\ell\neq\ell_{0}\) gives \(c_j+\beta_{j_\ell}^\star=\mathbb E(Y\mid A_j=\ell)\). Therefore
\[
\beta_{j_\ell}^\star
=
\mathbb E(Y\mid A_j=\ell)-\mathbb E(Y\mid A_j=\ell_{0})
=
\tau_{j_\ell},
\]
where the last equality is the identification result of Section~\ref{sec:amce}, which holds under the identifying assumptions together with complete independent randomization (\citealp{hainmueller_causal_2014}, Proposition~3).
\end{proof}

\paragraph{Beyond complete independent randomization.}
Complete independent randomization ensures that the indicator blocks for different attributes are orthogonal, which is why each AMCE equals the corresponding coefficient from the additive main-effects regression. Under restricted randomization, the attributes involved in the restrictions may be dependent, so a coefficient from this simple regression need not equal the AMCE even though the AMCE remains identified under the appropriate conditional randomization assumptions \citep{hainmueller_causal_2014}.

In that setting, the population regression can be expanded to account for the attributes involved in the restrictions. The AMCE is then recovered by calculating the relevant coefficient contrasts and averaging them using weights determined by the conjoint design. It therefore remains a linear contrast \(R\beta^\star\) from an appropriately specified population least-squares regression and is still within the scope of Theorem~\ref{thm:cluster_robust_e_process}.

\section{Proof of Theorem~\ref{thm:conjoint_estimands}}
\label{appendix:conjoint_theorem}

\begin{proof}
The population least-squares first-order condition gives \(\mathbb E(\psi_i)=0\). Because cluster sizes, regressor entries, and outcomes are uniformly bounded, \(X_i'X_i\) and \(\psi_i=X_i'(Y_i-X_i\beta^\star)\) are bounded. Consequently,
\[
\mathbb E\|X_i'X_i\|^2<\infty,
\qquad
\mathbb E\|\psi_i\|^s<\infty
\quad\text{for some }s>3.
\]
Together with the i.i.d.\ respondent clusters and the assumed positive definiteness of \(H\) and \(\Sigma_R\), these are the primitive conditions of \citet[Lemma~2]{lindon_clustered_2026}. That lemma verifies the cluster-score regularity required by \citet[Theorem~2]{lindon_clustered_2026}, restated as Theorem~\ref{thm:cluster_robust_e_process} here. The asymptotic \(e\)-process follows, and Corollaries~\ref{cor:av_pvalue}--\ref{cor:cluster_cs} give the asymptotic anytime-valid \(p\)-values and confidence sequence.

For AMCEs, Proposition~\ref{prop:amce_projection} identifies each AMCE with its population least-squares coefficient, so any stated contrast of the AMCEs has the required representation \(R\beta^\star\). For other conjoint estimands, the same conclusion applies once their identification and representation as a coefficient or linear contrast of population OLS coefficients have been established. No linear conditional-mean model or independence of outcomes within a respondent is required.
\end{proof}

\section{\texorpdfstring{Conventional fixed-$G$ cluster-robust inference}{Conventional fixed-G cluster-robust inference}}
\label{appendix:conventional_cluster_inference}

Let \(N_G=\sum_{g=1}^Gn_g\) denote the total number of observation-level rows and
recall from Section~\ref{sec:fixed_g_reference} that
\(\widehat s_g=X_g'\widehat u_g\) is cluster \(g\)'s estimated score contribution. The uncorrected one-way cluster-robust covariance estimator, commonly called CR0, is
\[
\widehat V_{\mathrm{CR0}}
=
A_G^{-1}
\left(
\sum_{g=1}^{G}X_g'\widehat u_g\widehat u_g'X_g
\right)
A_G^{-1},
\]
and the estimator \(\widehat V_G=\frac{G}{G-1}\widehat V_{\mathrm{CR0}}\), used in
Theorem~\ref{thm:cluster_robust_e_process}, applies the finite-cluster correction
\(G/(G-1)\) (CR1). Under standard regularity conditions for independent clusters,
\[
\sqrt{G}(\widehat\beta_G-\beta^\star)
\rightsquigarrow
N\!\left(0,H^{-1}\Omega H^{-1}\right),
\]
and \(\widehat V_{\mathrm{CR0}}\), hence \(\widehat V_G\), consistently estimates the variance of
\(\widehat\beta_G\) itself, namely \(H^{-1}\Omega H^{-1}/G\)
\citep{colin_cameron_practitioners_2015, mackinnon_cluster-robust_2023}. This permits arbitrary dependence and
heterogeneous variances within clusters while relying on independence across
clusters.

For \(q=1\), so that \(R\) is a row vector and
\(\theta=R\beta^\star\) is scalar, the conventional statistic is
\[
T_{\mathrm{CR}}
=
\frac{R\widehat\beta_G-\theta_0}
{\sqrt{R\widehat V_GR'}}.
\]
Under large-\(G\) asymptotics, this statistic has a standard normal reference
distribution, yielding the usual normal-quantile CI. With finitely many clusters,
software instead compares it with \(t_{G-1}\), giving the CI
\[
R\widehat\beta_G
\ \pm\ t_{G-1,1-\alpha/2}
\sqrt{R\widehat V_GR'}.
\]
The exact finite-cluster multiplier varies across software. The factor \(G/(G-1)\)
used here is also the SAS convention; Stata's \texttt{regress} and the default in
\texttt{fixest} use \(\frac{G}{G-1}\frac{N_G-1}{N_G-p}\)
\citep{colin_cameron_practitioners_2015}, whose additional factor converges to one as
\(N_G\to\infty\).

The \(t_{G-1}\) heuristic follows from the cluster-score linearization. If the
\(G\) independent score contributions behaved like equally weighted Gaussian
observations with common variance, estimating their dispersion from those same
clusters would yield the familiar one-sample \(t\) distribution with \(G-1\)
degrees of freedom. This explains why the cluster count, rather than \(N_G\), governs
the reference distribution. It does not make the approximation exact for a general
clustered regression.

The multi-parameter case is more delicate. For \(q\) restrictions
\(R\beta^\star=\theta_0\), define
\[
W_{\mathrm{CR}}
=
(R\widehat\beta_G-\theta_0)'
\left(R\widehat V_GR'\right)^{-1}
(R\widehat\beta_G-\theta_0).
\]
Under large-\(G\) asymptotics with fixed \(q\), \(W_{\mathrm{CR}}\) is commonly
compared with \(\chi_q^2\); as a finite-cluster approximation, we instead compare
\(W_{\mathrm{CR}}/q\) with \(F(q,G-q)\), which is the stricter of the two, since
\(\chi_q^2/q\) is the limiting \(F(q,\infty)\) law, while software such as Stata's
\texttt{regress} uses \(F(q,G-1)\), extending the \(t_{G-1}\) rule to joint tests
without adjustment for \(q\) \citep{colin_cameron_practitioners_2015}. There is no general exact finite-sample \(F\) law for a
cluster-robust Wald test. Under stronger conditions, as cluster sizes grow with
\(G\) fixed, \citet{bester_inference_2011} show that a scaled CR1 Wald statistic
has the limit
\[
\frac{G-q}{q(G-1)}W_{\mathrm{CR}}
\rightsquigarrow
F(q,G-q).
\]
Equivalently, the critical value for \(W_{\mathrm{CR}}\) is scaled by
\(q(G-1)/(G-q)\); the comparison of \(W_{\mathrm{CR}}/q\) with \(F(q,G-q)\) used in
the main text omits the factor \((G-1)/(G-q)=1+(q-1)/(G-q)\), which equals one when
\(q=1\). As \citet{bester_inference_2011} and
\citet{mackinnon_cluster-robust_2023} emphasize, this result relies on strong
assumptions, including fixed \(G\), equal cluster sizes, and limited within-cluster
dependence.

In conjoint regressions, clusters may differ in size and leverage, score variances may
be heterogeneous, and nuisance parameters may affect the statistic. Multi-restriction
tests are especially fragile because the cluster-robust covariance matrix has rank at
most \(G-1\), since the estimated cluster scores sum to zero by the normal equations, making its inverse unstable when \(q\) is not
small relative to \(G\) \citep{mackinnon_cluster-robust_2023}. Bias-reduced CR2
estimators, Satterthwaite or Hotelling \(T^2\) approximations, and wild cluster
bootstrap methods are often preferable when \(G\) is small or cluster leverage is
uneven \citep{pustejovsky_small-sample_2016,
mackinnon_cluster-robust_2023}.

Like conventional cluster-robust inference, the anytime-valid construction applied
in this paper relies on cluster-level asymptotics and does not provide an exact
finite-sample law for arbitrary clustered regressions.

\section{Simulation details}
\label{appendix:sim_details}

Unless stated otherwise, the empirical regressions use the default finite-cluster
correction implemented by \texttt{fixest}, which multiplies the CR1 covariance
\(\widehat V_G\) of Section~\ref{sec:fixed_g_reference} by the additional factor
\((N_G-1)/(N_G-p)=1+(p-1)/(N_G-p)\), where \(N_G\) is the number of profile-level rows
and \(p\) the number of regressors. This factor converges to one as \(N_G\to\infty\)
and is asymptotically immaterial.

\subsection{Illustrative example}
\label{appendix:illustrative_example}

As noted in Section~\ref{sec:illustrative_example}, the candidate profiles are drawn from a joint uniform distribution (profile attributes are independently and uniformly randomized). The simulated profile choices are intentionally not generated from the additive linear probability model used for estimation. Instead, simulated choices arise from a nonlinear random-utility model with a logistic choice rule. Let \(x=(p,r)\) denote a profile's party and region. We first define an index
\[
s(x)
=
b_p(p)+b_r(r)+b_{pr}(p,r),
\]
where the baseline coefficients are zero and
\[
b_p(\text{Left})=0.35,\qquad
b_r(\text{South})=-0.06,\qquad
b_r(\text{East})=-0.27,\qquad
b_r(\text{West})=0.16.
\]
We also include party-region interactions
\[
b_{pr}(\text{Left},\text{South})=0.04,\qquad
b_{pr}(\text{Left},\text{East})=-0.05,\qquad
b_{pr}(\text{Left},\text{West})=0.03,
\]
with all remaining interactions set to zero. Each respondent has a discrete taste parameter \(U_i\in\{-0.45,0,0.45\}\) with probabilities \((0.25,0.50,0.25)\). Let \(z(r)\) denote the centered region score, taking values \(-0.5,-1/6,1/6,0.5\) for North, South, East, and West, respectively. The respondent-specific latent utility of profile \(x=(p,r)\) is
\[
\eta_i(x)
=
s(x)
+0.35s(x)^2
-0.10s(x)^3
+0.12s(x)\mathbf{1}\{p=\text{Left}\}
+U_i\mathbf{1}\{p=\text{Left}\}
+0.08U_i z(r).
\]
This utility index is nonlinear in the randomized attributes for two reasons. First, \(s(x)\) enters through squared and cubic transformations. Second, the term \(0.12s(x)\mathbf{1}\{p=\text{Left}\}\) and the party-region coefficients \(b_{pr}(p,r)\) create interactions among profile attributes. In addition, \(U_i\) creates respondent-level preference heterogeneity and serial dependence across the profiles evaluated by the same respondent.

For each forced-choice task with profiles \(x_a\) and \(x_b\), the probability that profile \(x_a\) is chosen is
\[
\Pr(Y_a=1\mid x_a,x_b,U_i)
=
\operatorname{logit}^{-1}\{\eta_i(x_a)-\eta_i(x_b)\}.
\]
For each task, we draw \(Y_a\) from the corresponding Bernoulli distribution and set \(Y_b=1-Y_a\). Conditional on the randomized profiles and \(U_i\), choices are drawn independently across tasks. Marginally, however, outcomes contributed by the same respondent are dependent through their shared value of \(U_i\), while the two profile-level outcomes within each task are mechanically dependent.
Thus the outcome is binary and the conditional mean is nonlinear in the randomized attributes, with interactions and respondent-level dependence across tasks. The OLS regression used in the main analysis is therefore simply an estimator of the AMCE, not a correctly specified conditional-mean model.

For any respondent type \(u\), let \(\eta_u(x)\) denote the latent utility of profile \(x\) obtained by evaluating \(\eta_i(x)\) at \(U_i=u\). Accordingly, \(\eta_U(x)\) denotes the same utility function evaluated at the random respondent type \(U\), distributed as specified above and independently of the randomized profiles. The exact AMCE for level \(\ell\) of attribute \(j\), relative to baseline \(\ell_{0}\), is
\[
\tau_{j_\ell}
=
\mathbb E_{A_{-j},\,X',\,U}
\left[
\operatorname{logit}^{-1}\{\eta_U(\ell,A_{-j})-\eta_U(X')\}
-
\operatorname{logit}^{-1}\{\eta_U(\ell_{0},A_{-j})-\eta_U(X')\}
\right],
\]
where \(X'\) is the competing profile drawn from the same randomized profile distribution. Because all attributes and respondent types are discrete, this expectation is evaluated by finite summation. The resulting exact AMCEs are
\[
\tau_{\text{South}}=-0.0103,\qquad
\tau_{\text{East}}=-0.0768,\qquad
\tau_{\text{West}}=0.0507,\qquad
\tau_{\text{Left}}=0.1001.
\]

\subsection{Naive peeking inflates false positives}
\label{appendix:false_positive_sim_details}

The false-positive simulations use the same forced-choice design as Section~\ref{sec:illustrative_example}. Candidate party and region are independently and uniformly randomized, and each respondent completes two tasks containing two profiles. To generate data under the null, we set all utility coefficients and respondent-level preference heterogeneity to zero. Each profile is therefore chosen with probability \(1/2\), independently of its attributes. For each task, we draw the choice outcome from a Bernoulli distribution with probability \(1/2\) and assign an outcome of one to the chosen profile and zero to the other profile. Consequently, all party and region AMCEs are exactly zero.

At each interim analysis, we regress the profile-level choice indicator on indicators for South, East, West, and Left, with North and Right serving as the respective reference levels. The covariance matrix is estimated by clustering all profile-level observations from the same respondent. For the scalar hypothesis
\(
H_0:\beta_{\text{South}}=0
\),
the conventional procedure uses the cluster-robust \(t\)-statistic and a \(t_{G-1}\) reference distribution. The anytime-valid procedure applies Theorem~\ref{thm:cluster_robust_e_process} with \(q=1\). For the joint hypothesis
\(
H_0:
\beta_{\text{South}}
=
\beta_{\text{East}}
=
\beta_{\text{West}}
=
0
\),
let \(Q_G\) denote the respondent-cluster-robust Wald statistic for the three restrictions. The conventional procedure compares \(Q_G/3\) with an \(F(3,G-3)\) reference distribution. The anytime-valid procedure transforms the same Wald statistic using Theorem~\ref{thm:cluster_robust_e_process} with \(q=3\). Each procedure rejects when its corresponding \(p\)-value falls below \(\alpha=0.05\).

For Figure~\ref{fig:type1_error}, we begin monitoring at \(G=100\), conduct an interim analysis after every 10 additional respondents through \(G=2{,}500\), and use 1,000 repeated simulations. Each respondent completes two tasks containing two profiles, so the final profile-level row count is \(N_G=10{,}000\). The anytime-valid tuning parameters are fixed before monitoring at the values optimized for the respondent-cluster horizon: \(\lambda=303.87\) for the scalar test and \(\lambda=712.03\) for the three-dimensional test.

At each interim sample size \(G\), the cumulative Type I error rate is estimated as the proportion of simulations in which the relevant procedure has rejected the true null at least once at or before \(G\). Thus, a simulation continues to count as a false positive at every subsequent value of \(G\) after its first erroneous rejection. Monte Carlo 95\% intervals are calculated from the sampling variation across repeated simulations.

\begin{figure}[!htbp]
    \centering
    \includegraphics[width=0.9\textwidth]{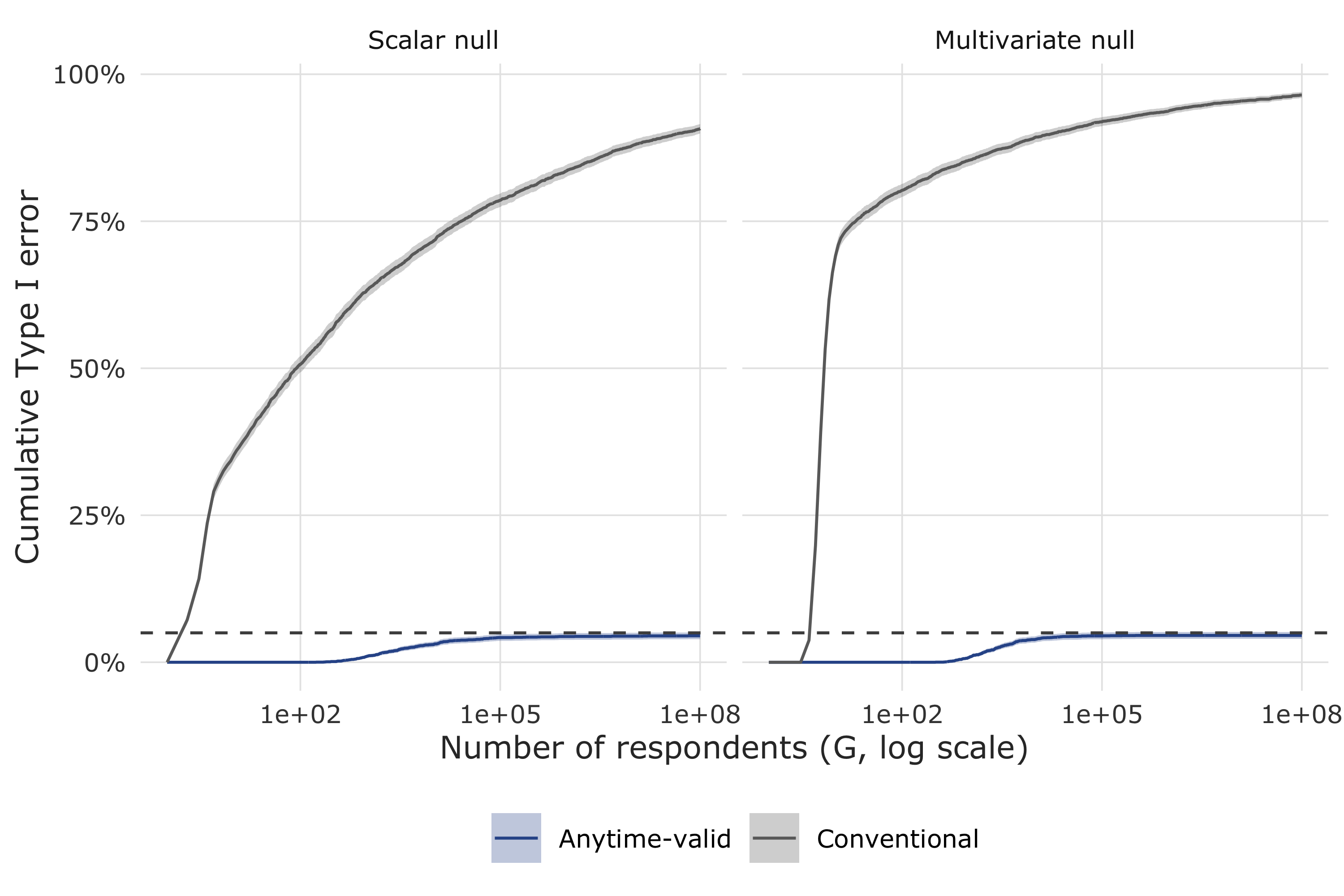}
    \caption{\textbf{Cumulative Type I error rates over a long monitoring horizon.} The figure reports the same scalar and joint null tests as Figure~\ref{fig:type1_error}, with sequential monitoring on a dense correlation-spaced grid extended through \(G=10^8\). The horizontal dashed line denotes the nominal Type I error level of \(\alpha=0.05\).}
    \label{fig:type1_error_long_horizon}
\end{figure}

For the long-horizon results in Figure~\ref{fig:type1_error_long_horizon}, we use 5,000 repetitions through \(G=10^8\). To make this horizon computationally feasible, each respondent cluster contributes one forced-choice task rather than the two tasks used in Figure~\ref{fig:type1_error}; party and region remain independently and uniformly randomized, and choices are generated under the same sharp null. We use a dense monitoring grid under which adjacent standardized null statistics have approximate correlation \(\rho=0.995\). Under the Brownian approximation, the grid satisfies
\(
G_{k+1}\approx\frac{G_k}{\rho^2}
\).
The grid includes integer-valued looks at the smallest estimable cluster counts and then uses the correlation-spaced sequence through \(G=10^8\), for 1,479 looks in total. We use the same tuning parameters as in Figure~\ref{fig:type1_error}, fixed before monitoring and optimized for the target horizon \(G^\star=2{,}500\): \(\lambda=303.87\) for the scalar test and \(\lambda=712.03\) for the three-dimensional test.

At \(G=10^8\), the cumulative false-positive rates under the anytime-valid procedure are below the nominal 5\% rate for both the scalar and the joint test. In contrast, the corresponding conventional false-positive rates are both above 90\%.

\subsection{Sample-efficiency gains under plausible conjoint parameters}
\label{appendix:sample_efficiency_full}

The sample-efficiency simulations use the same nonlinear random-utility model as
Appendix~\ref{appendix:illustrative_example}. Two attributes are randomized independently and
uniformly: \emph{Party}, with levels Right (baseline) and Left, and \emph{Region},
with \(n_{\text{levels}}\) levels. Each respondent completes two tasks of two profiles, so
\(n_i=4\). The Party coefficient is zero. Region coefficients are calibrated so that the first level has the target AMCE, with the remaining levels spanning a total range of \(0.01\).

The grid evaluates target AMCEs from \(0.02\) to \(0.12\) in steps of \(0.01\);
effective sample size caps \(N_{\max}\in\{3000,6000,11000,18000\}\); and
\(n_{\text{levels}}\in\{4,6,9\}\). Each unique \(\text{AMCE} \times N_{\max} \times n_{\text{levels}}\) combination is replicated 500 times. Within a
replication, the analyst refits the AMCE regression after every 100 respondents,
estimating by regressing the profile-level choice indicator on Party and Region
indicators with cluster-robust standard errors, and forms the scalar
confidence sequence of Corollary~\ref{cor:cluster_cs} with \(q=1\).

We use a significance-based stopping decision, as before. Data collection
stops when the confidence sequence for a focal AMCE excludes zero, and otherwise continues to the respondent horizon \(G_{\max}=N_{\max}/2\). Although continuous monitoring is indexed by respondents \(G\), we report effective sample size \(N_{\max}\) to match common conjoint practice. Because each respondent completes two tasks, \(N_{\max}=2G\), so both summary quantities are identical whether sample size is expressed in terms of respondents or effective sample size.

The tuning parameter \(\lambda\) is chosen, separately for each \(N_{\max}\), as the
value minimizing the half-width \(\sqrt{\kappa_G(\lambda,1,\alpha)}\) of the
confidence sequence evaluated at the respondent horizon \(G=G_{\max}=N_{\max}/2\),
giving \(\lambda=182.10\), \(364.76\), \(669.19\), and \(1095.40\) for
\(N_{\max}=3{,}000\), \(6{,}000\), \(11{,}000\), and \(18{,}000\). It is computed once,
before any data are observed, and held fixed at every interim analysis.

\begin{figure}[!htbp]
    \centering
    \includegraphics[width=1\textwidth]{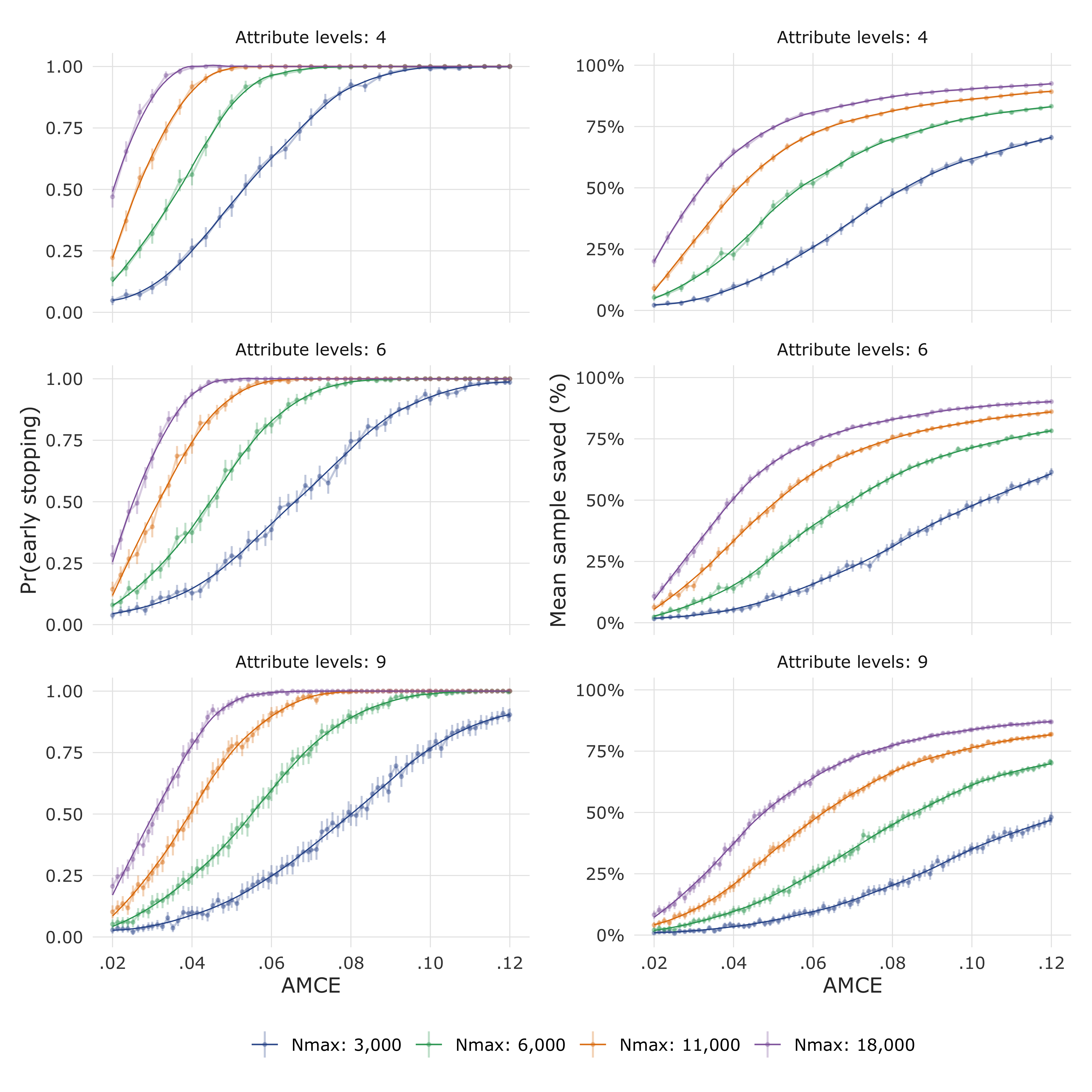}
    \caption{\textbf{Sample-efficiency gains under anytime-valid stopping across conjoint complexity.} The left column shows the probability of early stopping, and the right column shows the corresponding mean sample savings. Rows correspond to conjoint designs with 4, 6, and 9 levels in the largest attribute. In both panels, the x-axis shows the true AMCE effect size; colors denote maximum effective sample sizes \(N_{\max}\).}
    \label{fig:early_stopping_full}
\end{figure}

Figure~\ref{fig:early_stopping_full} reports the full version of the sample-efficiency results from Section~\ref{sec:sample_efficiency}, displaying outcomes for \(n_{\text{levels}} \in \{4,6,9\}\). The main text focuses on \(n_{\text{levels}}=6\), the median number of levels reported by \citet{stefanelli_subjects_2020}. Figure~\ref{fig:early_stopping_full} shows that the same qualitative pattern holds more broadly: the probability of early stopping and the corresponding mean sample savings both increase with the true AMCE and decline as the number of levels in the largest attribute grows.

\subsection{Sample-efficiency gains when the true effect exceeds the target}
\label{appendix:overshoot_details}

The overshoot simulations use the data-generating process, design, estimation, and
stopping rule of Appendix~\ref{appendix:sample_efficiency_full}, with
\(n_{\text{levels}}=6\) and 1{,}000 replications. The powered-for target AMCE is \(0.03\).

The fixed-sample benchmark \(G_{\text{fix}}\) is the number of respondents required to
detect the target AMCE of \(0.03\) with 80\% power at \(\alpha=0.05\) in a conjoint
whose largest attribute has six levels, obtained from the power formulae of
\citet{schuessler_power_2020}; their expression returns an effective sample size (number of respondents \(G\) \(\times\) number of tasks), which we convert to respondents. \(\lambda\) is calibrated to
\(G_{\text{fix}}\). Monitoring continues to \(5G_{\text{fix}}\) to ensure an accurate representation of sample savings, for example capturing runs where the anytime-valid approach provides less sample efficiency than the conventional approach. Reported savings are
\(1-G_{\text{stop}}/G_{\text{fix}}\), averaged over replications.

Calibrating \(\lambda\) to \(G_{\text{fix}}\) makes the reported savings conservative
at the larger ratios. A study whose true effect is three times the target stops near
\(G=1{,}300\), where the boundary implied by \(\lambda=795\) is about 10\% wider than
one calibrated to that sample size; at ratios \(1\) and \(1.25\) the discrepancy is
under \(1\%\). Thus, the reported gains at large ratios should be read as a lower bound.

\subsection{Asylum-seeker conjoint case study}
\label{appendix:asylum_case_details}

The case study uses the 2016 wave of the asylum-seeker conjoint experiment originally analyzed by \citet{bansak_how_2016} and released in the replication archive of \citet{bansak_replication_2023}. The raw public file contains 18,030 respondents and 180,300 profile-level observations. We remove nine respondents for whom all profile attributes are missing, leaving \(G=18{,}021\) respondents and 180,210 profile-level observations.

Each respondent completes five tasks containing two profiles. The profile-level outcome equals one when the profile is selected for admission and zero otherwise. We regress this outcome on dummy variables for the non-baseline levels of the nine randomized attributes: testimony consistency, gender, country of origin, age, previous occupation, vulnerability, reason for migrating, religion, and language skills. The resulting 27 non-intercept coefficients are AMCE estimates relative to the corresponding baseline levels. We estimate the cluster-robust covariance matrix.

The sequential process is indexed by respondents. Because fielding timestamps are unavailable, respondents enter the monitoring process in the order they appear in the public data file. We refit the AMCE regression after every 100 respondents and always include the final fit at \(G=18{,}021\). Consequently, the exercise illustrates one of potentially many ways the data could have been sequentially collected.

We use \(\alpha=0.05\) for the conventional confidence intervals, anytime-valid CSs, and \(e\)-process rejection thresholds. The anytime-valid tuning parameter is fixed throughout monitoring at \(\lambda=2193.92\), the value optimized for the planned horizon \(G=18{,}021\).

Figure~\ref{fig:asylum_eprocess_growth} displays one AMCE from each randomized attribute. For visualization, we select the non-baseline level having the largest absolute final cluster-robust \(t\)-statistic within that attribute. This selection is used only to determine which trajectories are displayed; all 27 AMCEs are estimated and analyzed using the same procedure. The \(e\)-processes are plotted on the natural-log scale to avoid numerical overflow and to make the rejection threshold \(\log(1/\alpha)=\log(20)\) visually comparable across panels.

Figure~\ref{fig:asylum_case_study} reports all 27 final AMCE estimates. The conventional intervals are fixed-sample respondent-clustered 95\% confidence intervals. The anytime-valid intervals are the final values of the respondent-clustered 95\% confidence sequences generated by the complete monitoring process.

\section{Anytime-valid coverage --- empirical evidence}
\label{appendix:coverage}

\begin{figure}[!htbp]
    \centering
    \includegraphics[width=0.8\textwidth]{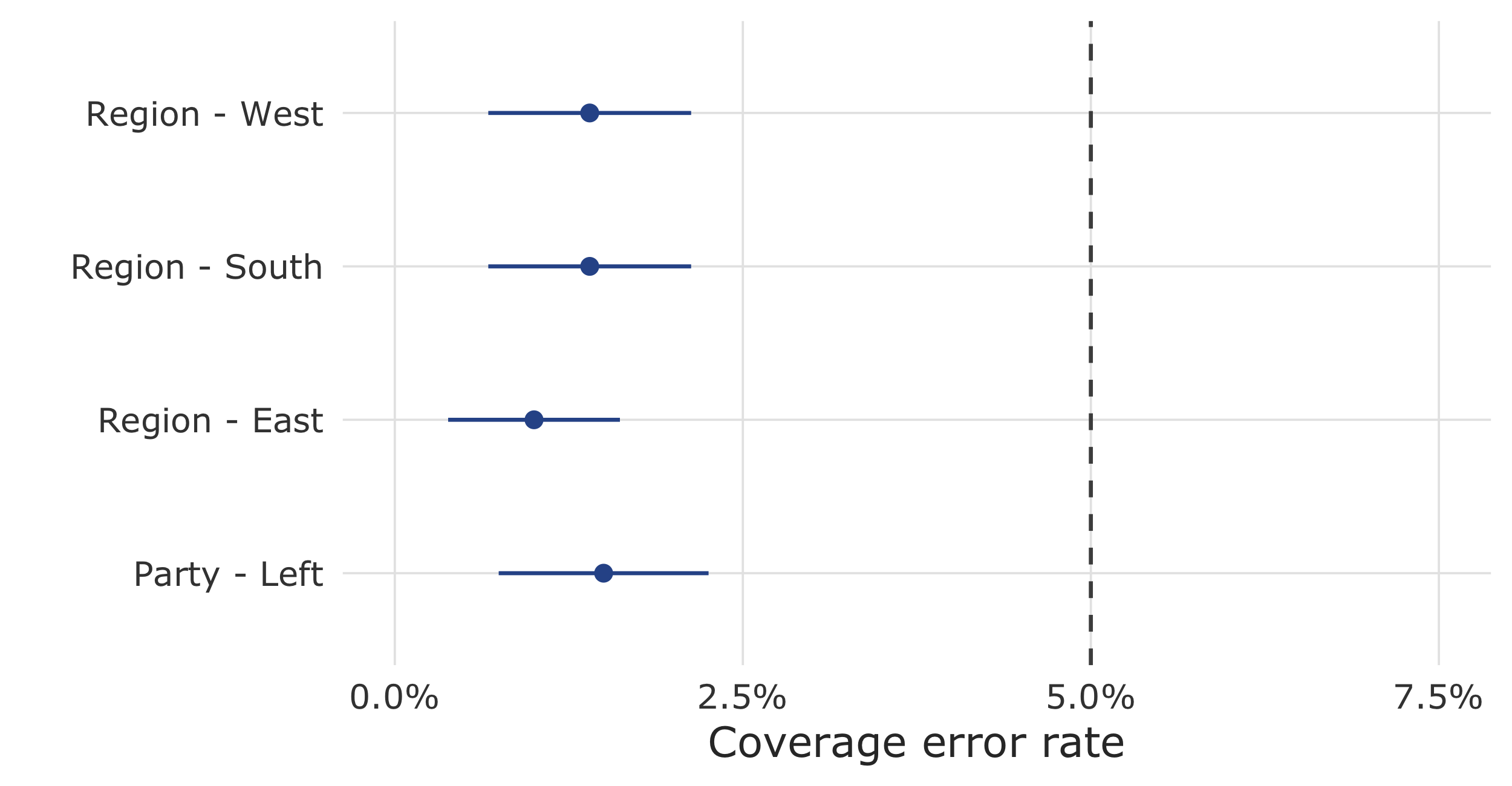}
    \caption{\textbf{AMCE coverage error rates.} The y-axis shows each attribute level for which an AMCE is estimated, and the x-axis shows the mean (\(\pm\) 95\% CIs) coverage error rate of 95\% confidence sequences across 1,000 simulations. The vertical dashed line represents the nominal coverage error level of \(\alpha = 0.05\).}
    \label{fig:amce_coverage}
\end{figure}

Following the simulation setup detailed in Appendix~\ref{appendix:illustrative_example}, we assess the finite-sample coverage of the asymptotic anytime-valid procedure under sequential monitoring. Figure~\ref{fig:amce_coverage} reports empirical coverage error rates for our 95\% CSs across 1,000 repeated simulations. For each AMCE, we define the coverage error rate as the proportion of simulations in which the CS fails to include the true parameter value at one or more of the interim analyses, conducted every 50 respondents through \(G=2{,}500\). This measures error over the monitoring sequence; the corresponding pointwise error rate at any single \(G\) is necessarily no larger. Across all four AMCEs, the observed error rates remain below the nominal level of \(\alpha = 0.05\).

\end{document}